 \date{1997.8.12}
 \documentstyle[12pt]{article}
 \newif\ifwithoutmath
 \ifwithoutmath
  \def\BBb#1{{\mathchoice
   {\mbox{\normalsize\bf #1}}
   {\mbox{\normalsize\bf #1}}
   {\mbox{\scriptsize\bf #1}}
   {\mbox{\tiny\bf #1}}}}
  \def\spt#1{{\mathchoice
   {\mbox{\normalsize\bf #1}}
   {\mbox{\normalsize\bf #1}}
   {\mbox{\scriptsize\bf #1}}
   {\mbox{\tiny\bf #1}}}}
 \else
  \font\ffx=msbm10  scaled \magstephalf
  \font\ffvii=msbm7 scaled \magstephalf
  \font\ffv=msbm5   scaled \magstephalf
  \newfam\msbfam    \textfont\msbfam=\ffx 
                    \scriptfont\msbfam=\ffvii
                    \scriptscriptfont\msbfam=\ffv
  \def\BBb#1{\fam\msbfam #1}%
  \font\fffx=eusm10 scaled \magstephalf
  \font\fffvii =eusm7  scaled \magstephalf
  \font\fffv =eusm5 scaled \magstephalf
  \newfam\eusfam    \textfont\eusfam=\fffx  
                    \scriptfont\eusfam=\fffvii
                    \scriptscriptfont\eusfam=\fffv
  \def\spt#1{{\fam\eusfam #1}}%
 \fi
 \newtheorem{prp}{Proposition}

 \catcode`@=11
 \@addtoreset{equation}{section}\@addtoreset{prp}{section}
 \def\@begintheorem#1#2{\it \trivlist \item[\hskip \labelsep{\bf #1\ #2.\ }]}
 \def\@opargbegintheorem#1#2#3{\it \trivlist
       \item[\hskip \labelsep{\bf #1\ #2\ (#3).}]}
 \catcode`@=12
 \def\appendix{\par\setcounter{section}{0}\setcounter{subsection}{0}
    \def\thesection{\Alph{section}}
    \vspace{3.5ex plus 1ex minus .2ex}\par\noindent{\Large\bf Appendix}
    \nopagebreak\vspace{2.0ex plus .2ex}\nopagebreak\\ \nopagebreak}
 \newtheorem{lem}[prp]{Lemma}
 \newtheorem{thm}[prp]{Theorem}
 \newtheorem{cor}[prp]{Corollary}
 \newenvironment{prf}{\begin{trivlist} \item[{\em Proof.}]}{\end{trivlist}
   \bigskip \par}
 \newenvironment{prfof}[1]{\begin{trivlist} \item[{\em Proof of #1.}]}{%
   \end{trivlist} \bigskip \par}
 \def\qed{\relax\ifmmode\let\@tempa\relax\ifcase\@eqcnt\def\@tempa{& & &}\or
   \def\@tempa{& &}\else\def\@tempa{&}\fi\@tempa $\Box$ \else\hfill 
   {\footnotesize$\Box$} \fi}
 \newenvironment{rem}{\begin{trivlist} \item[{\em Remark.}]}{\ignorespaces
   \end{trivlist} \bigskip \par}
 \def\prpb{\begin{prp}}\def\prpe{\end{prp}}
 \def\lemb{\begin{lem}}\def\leme{\end{lem}}
 \def\thmb{\begin{thm}}\def\thme{\end{thm}}
 \def\corb{\begin{cor}}\def\core{\end{cor}}
 \def\prfb{\begin{prf}}
 \def\prfe{\hfill{\footnotesize$\Box$}\end{prf}}
 \def\prfofb#1{\begin{prfof}{#1}}
 \def\prfofe{\hfill{\footnotesize$\Box$}\end{prfof}}
 \def\remb{\begin{rem}}\def\reme{\end{rem}}
 \def\itmb{\begin{itemize}}\def\itme{\end{itemize}}
 \def\prpa#1{\label{p:#1}}\def\prpu#1{Proposition~\ref{p:#1}}
 \def\lema#1{\label{l:#1}}\def\lemu#1{Lemma~\ref{l:#1}}
 \def\thma#1{\label{t:#1}}\def\thmu#1{Theorem~\ref{t:#1}}
 
 \def\seca#1{\label{s:#1}}\def\secu#1{Section~\ref{s:#1}}
 \def\parr{\par\noindent}
 \def\bparr{\bigskip\par\noindent}  
 \def\mparr{\medskip\par\noindent}  
 \def\eqb{\begin{eqnarray}}         \def\eqe{\end{eqnarray}}
 \def\eqsb{\begin{eqnarray*}}       \def\eqse{\end{eqnarray*}}
 \def\eqa#1{\label{e:#1}}           \def\equ#1{(\ref{e:#1})}
 \def\integer{{\BBb Z}}
 \def\complex{{\BBb C}}
 \def\real{{\BBb R}}
 
 \def\T{{\BBb T}}
 \def\V{{\BBb V}}
 \def\O{{\spt O}}
 
 \def\const{{\it const.}}
 \def\dfrac{\displaystyle\frac}
 
 \def\Tr{{\rm Tr\,}}
 \def\det{{\rm det\,}}
 
 \def\ad{{\rm ad}}
 \def\norm#1{\parallel #1\parallel}
 \def\dV{\V^d_a}
 \def\tV{\dot\V^d_a}
 \def\torus{{\T^d_a}}
 \def\f{\complex^{2^{d/2}}}
 
 \def\dist#1{\langle{#1}\rangle}
 
 \def\Slash#1{
  \begin{picture}(5,6)(0,0)
  \put(-.7,-1.2){\line(5,6)6}
  \end{picture}
  \kern-.8em#1}
 \def\slash#1{
  \begin{picture}(5,6)(0,0)
  \put(-1.5,-1.7){\line(5,6)5}
  \end{picture}
  \kern-.8em#1}
 
\begin{document}
 \title
   {Mathematical Derivation of Chiral Anomaly \\
    in Lattice Gauge Theory with Wilson's Action}
 \author  {
           Tetsuya Hattori
   \\ {\small
           Department of Mathematics, Faculty of Science, Rikkyo University,
 } \\ {\small
           Nishi-Ikebukuro, Tokyo 171, Japan
 } \\ {\small e-mail address:\tt\ 
           hattori@rkmath.rikkyo.ac.jp
 } \\ \and
           Hiroshi Watanabe
   \\ {\small
           Department of Mathematics, Nippon Medical School,
 } \\ {\small
           2-297-2, Kosugi, Nakahara, Kawasaki 211, Japan
 } \\ {\small e-mail address:\tt\ 
           d34335@m-unix.cc.u-tokyo.ac.jp
 }        }
 \maketitle
 \begin{abstract} \noindent
  Chiral $U(1)$ anomaly is derived with mathematical rigor
  for a Euclidean fermion coupled to a smooth external $U(1)$ gauge field
  on an even dimensional torus
  as a continuum limit of lattice regularized fermion field theory
  with the Wilson term in the action.
  The present work rigorously proves for the first time that 
  the Wilson term correctly reproduces the chiral anomaly.
 \end{abstract}
 \mparr
 {\bf Running title:} Mathematical Derivation of Chiral Anomaly
 \section{Introduction}
 It is widely believed that continuum limit of
 a lattice regularized theory of quantized fermion 
 gives the correct chiral anomaly \cite{A,B,BJ}
 if we have Wilson terms
 in the lattice fermion action \cite{W}.
 Since Karsten and Smit \cite{KS} observed this fact 
 by a perturbative argument,
 there appeared mathematically more careful analyses \cite{K,SS},
 in which it was claimed that 
 the Wilson fermion has the expected continuum limit 
 and gives the correct chiral anomaly
 under some mathematical ansatz on perturbative expansions.
 However, a further investigation on the validity of the ansatz
 has not been published so far,
 and from a mathematical view point,
 a proof that Wilson's formalism for lattice fermion is a correct scheme
 giving the expected anomaly, has not been completed.

 On the other hand, in modern physics, the Wilson fermion provides 
 not only a mathematical basis for analytic investigations
 but also a practical scheme for numerical studies
 on Euclidean quantum field theories including fermions.
 In view of this,
 we cannot help asking for a mathematically rigorous proof
 that the Wilson term correctly reproduces
 the chiral anomaly in the continuum limit.
 
 In the present paper, 
 we study Wilson's formalism for a Euclidean lattice fermion
 coupled to a smooth external $U(1)$ gauge field
 defined on an even dimensional torus
 and derive the expected chiral anomaly in the continuum limit
 with mathematical rigor.
 The present work rigorously proves for the first time that
 the Wilson term correctly reproduces the chiral anomaly.
 
 We summarize below three essential mathematical problems in the present study
 and the strategies to solve the problems.
 \mparr
 {\it 1. Chiral oscillation.\ }
 Let $\torus=(a\integer/L\integer)^d$ be a discrete torus
 with period $L$ and lattice spacing $a$,
 where $L/a$ is a positive integer.
 We denote by $\dV$ the vector space consisting of all functions
 $u\,:\,\torus \longrightarrow \f$
 defined on $\torus$ with values in the `fibre' $\f$.
 The vacuum expectation of divergence of chiral current, 
 which is our main quantity of interest,
 is written in the form of $\Tr(K\gamma_{d+1})$,
 where $K$ is an operator acting on $\dV$,
 and $\gamma_{d+1}$ is the chirality in $d$ dimensions (see \equ{H0H1}).
 [Note that $\gamma_{d+1}$ is regarded as 
 an operator acting on $\dV$ instead of the fibre $\f$
 and therefore the multiplicities of the eigenvalues $\pm1$ 
 are not uniformly bounded in the lattice spacing $a$.]

 To take the trace, we choose,
 as in \cite{K,SS}, the `planewave basis', i.e.
 the set of eigenfunctions of the (free) translations on the lattice
 (see \equ{planewavebasis}).
 Though this choice may be standard, it should be underlined
 because the Wilson term is too weak
 to make the {\em trace norm} of $K\gamma_{d+1}$ uniformly bounded
 in the lattice spacing $a$.
 In other words, 
 it is essential to `cancel' (see e.g. \equ{zjto0})
 the highly degenerate positive and negative
 eigenvalues of $\gamma_{d+1}$ (the `chiral oscillation')
 before taking the continuum limit.
 The planewave basis is convenient for this purpose.
 \mparr
 {\bf Strategy 1:} {\it 
  Take the trace with respect to the planewave basis
  in order to explicitly cancel the chiral oscillation.}

 It may be illustrative to compare the situation with a previous
 related work by one of the authors \cite{Wat,WYPJA,WYLMP}.
 There, the chiral oscillation is controlled by introducing
 an additional heat-kernel regularization,
 with which the operator in question lies in the trace class,
 and the continuum limit can be taken without explicit cancellation
 of chiral oscillation.
 As a result, the index of the continuum Dirac operator appears, 
 which is equal to the Chern class
 by the index theorem \cite{ABP}, hence the proof is completed.
 \mparr
 {\it 2. Perturbative expansion. }
 After performing the fermion integration, we have a formula with inverse
 of operators (see \equ{firsttrace}).
 If we try to expand the inverse operators perturbatively,
 the convergence of the expansion becomes a problem,
 as is usual with formal perturbation series.
 For example, the operator $C$ defined in \equ{C} is decomposed as
 \eqb
  C &=& C_0+C_1
 \eqe
 into a {\em free part} $C_0$ (for vanishing gauge potentials)
 and an {\em interaction part} $C_1$,
 but the Neumann series
 \eqb\eqa{neumannC}
  C^{-1}&=&\sum_{\ell=0}^{\infty}(-C_0^{-1}C_1)^\ell C_0^{-1}
 \eqe
 is not absolutely convergent,
 unless the gauge field is sufficiently small.
 The following idea was suggeted in \cite{SS} 
 without details:
 \mparr
 {\bf Strategy 2:} {\it 
  Terminate a perturbative expansion of inverse operator
  up to a finite order, and use `positivity' to control
 the remainder terms.}
 \mparr
 Namely, we use an identity
 \eqb\eqa{finiteexp}
  C^{-1}&=&C_0^{-1}+\sum_{j=1}^{m-1}(-C_0^{-1}C_1)^jC_0^{-1}
           +(-C_0^{-1}C_1)^mC^{-1}\ ,\quad m=1,2,\dots\ ,
 \eqe
 for a sufficiently large $m$.
 What we need is the power-counting property
 of the operator $(-C_0^{-1}C_1)^j$,
 and an {\it a priori bound} arising from positivity of $C$,
 to control the remainder terms.
 See \secu{ndo} and \secu{order} for details.
 It may be suggestive to note that
 remainder estimates based on {\it positivity}
 is used in the convergence proof of cluster expansions
 in constructive field theory.
 \mparr
 {\it 3. Power counting.}
 The strategy 2 should be accompanied by a rigorous power-counting argument.
 Furthermore, we need asymptotic estimates of operators,
 when we take the continuum limit.
 Our last point is 
 to employ the power-counting arguments with mathematical rigor
 in order to extract the asymptotic behaviors.
  \mparr
 {\bf Strategy 3:} {\it Show and use the fact that 
 the operators in question are `quasi diagonal'
 with respect to the planewave basis.
 }
 \mparr
 The class of {\it quasi diagonal operators} defined in \secu{ndo}
 is closed with respect to the summation and the product,
 and, under a simple assumption,
 it is also closed with respect to the inverse operation.
 This calculus,
 which is a systematic arrangement
 of the asymptotic analysis of lattice operators
 in the spirit of power-counting,
 fits our analysis in this paper well.
 The smoothness assumption on the gauge fields is of relevance here.
 See \secu{ndo} for details.
 \mparr

 According to the above strategies,
 we take the continuum limit of $\Tr(K\gamma_{d+1})$
 and confirm that the Wilson fermion gives
 the correct chiral anomaly.

 This paper is organized as follows.
 We describe the model and 
 state the result (\thmu{maintheorem}) in Section 2.
 \thmu{maintheorem} is proved in the subsequent sections.
 In Section 3, we derive the expression of chiral anomaly
 in the form of chiral oscillation.
 Section 4 is devoted to the calculus of quasi diagonal operators.
 In Section 5, we apply the framework of Section 4 to several operators
 and determine their orders.
 The proof of \thmu{maintheorem} is completed in Section 6.

 The propositions in Section 3 and Section 6 may guide the
 readers who wish to have an overview of our argument.
 
 \section{Problem and Result}\seca{result}
 In this section,
 we formulate the problem and then state the result in \thmu{maintheorem}.
 \subsection{Notation}
 Let $\torus$ be a discrete torus with lattice spacing $a$ and period $L$.
 Throught this paper, we assume that $0<a<1/2, L/a\in\integer$.
 Let $\dV$ be the set of all mappings
 \eqb
  u=(u_s)_{s=1,2,\dots,2^{d/2}}\!\!\! &:& 
  \torus \longrightarrow \complex^{2^{d/2}}.
 \eqe
 We impose
 the anti-periodic boundary condition for definiteness,
 though boundary conditions play no essential roles in this paper.\footnote{%
 The anti-periodic boundary condition is required
 in the proof of the reflection positivity \cite{S}.}

 Let $(\ ,\ )$ be the inner product on $\dV$ defined by
 \eqb\eqa{innerproduct}
  (v,u)&=&a^d\sum_{x\in\torus}(v(x),u(x))_x\,,
 \eqe
 where $(\ ,\ )_x$ is the (standard) inner product on
 $\f$:
 \eqb  \nonumber
  (v(x),u(x))_x  &=&\sum_{s=1}^{2^{d/2}}\bar{v}_s(x)u_s(x)\,.
 \eqe

 Let $A_\mu(x), \mu=1,2,\dots,d,$ be a fixed smooth $U(1)$ gauge field
 on the continuum torus $\T^d=(\real/L\integer)^d$.
 Using $A_\mu$, we assign the gauge group element
 \eqb\eqa{U}
  U_{\mu}(x)&=&\exp\left(-iQ \int_0^a A_{\mu}(x+se_\mu)\,ds\right)
 \eqe
 to a directed link 
 $(x,x+e_{\mu}), x\in\torus, \mu=1,2,\dots,d$,
 where $Q\in\real$ is a charge and 
 $e_\mu$ is a unit vector along the $\mu$-th axis for $\mu=1,2,\dots, d$.
 Denote the free and the covariant translations by $T_{\mu0}$ and $T_{\mu}$,
 respectively:
 \eqb\eqa{freetranslation}
  T_{\mu0}u(x)&=&u(x+ae_\mu) ,\ \mu=1,2,\dots,d,\ x\in\torus\,,\ u\in\dV\,,\\
  T_{\mu}&=&U_{\mu}T_{\mu0} \,,\ \mu=1,2,\dots,d,   \eqa{covtranslation}
 \eqe
 where $U_\mu$ denotes the multiplication operator defined by
 \eqb
  U_\mu u(x)&=&U_\mu(x)u(x),\ \mu=1,2,\dots,d,\ x\in\torus\,,\ u\in\dV\,.
 \eqe
 Note that $T_{\mu}^*=T_{\mu}^{-1}$ holds,
 where $T_{\mu}^*$ is the adjoint with respect to 
 the inner product $(\ ,\ )$.
 Put
 \eqb\eqa{Dmu}
  D_\mu&=&\dfrac{1}{2a}(T_\mu-T_\mu^*)\,,\ \mu=1,2,\dots, d,
 \eqe
 and let $\gamma_\mu, \mu=1,2,\dots,d,$ be the $d$ dimensional 
 anti-hermitian Dirac matrices.
 Namely,
 \eqb
  \gamma_\mu\gamma_\nu + \gamma_\nu\gamma_\mu &=& 
    -2\delta_{\mu\nu}I_{2^{d/2}}  \ ,\quad\mu,\nu=1,2,\dots,d,
 \eqe
 where $I_\ell$ denotes the unit matrix of order $\ell$.
 In what follows, we abbreviate $\gamma_{\mu}\otimes\dot I$ to $\gamma_{\mu}$,
 where $\dot I$ denotes the identity operator on the space
 $\tV=\{\varphi\ :\ \torus\to\complex\}$.
 Then, 
 the lattice Dirac operator on the discrete torus $\torus$ is by definition
 \eqb\eqa{Diracoperator}
  \Slash D &=& \sum_{\mu=1}^{d}\gamma_\mu D_\mu\,.
 \eqe
 \subsection{Lattice fermion}
 As is well-known, the operator $\Slash D$ has eigenfunctions
 called {\it doublers}
 which do not approximate any eigenfunctions of 
 the continuum Dirac operator.
 In order to suppress the contributions of doublers 
 to vacuum expectation values,
 we introduce the Wilson term:
 \eqb\eqa{Wilson}
   W&=&-\sum_{\mu=1}^{d}(2I-T_\mu-T_\mu^*)\,,
 \eqe
 where $I$ denotes the identity operator on $\dV$.
 Then our lattice fermion action is written as 
 \eqb
  S(\bar\psi,\psi)&=&(\bar\psi,(i\Slash D +MI -\dfrac{r}{2a}W)\psi),
 \eqe
 where $M>0$ is the fermion mass and $r$ is a positive constant.\footnote{%
 Our analysis allows any $r>0$, but 
 the proof of reflection positivity \cite{S} requires $0<r\le1$.}
 Using the action $S$, we define the vacuum expectation by
 \eqb
  \langle \Phi(\bar\psi,\psi)\rangle
  &=&\dfrac{1}{Z(A)}\int \prod_{x\in\torus}d\bar\psi(x)d\psi(x)
     \exp(-S(\bar\psi,\psi))\Phi(\bar\psi,\psi),
 \eqe
 where the integrations with respect to $\bar\psi$ and $\psi$ are
 the Grassmann integrations \cite{B,S} and the partition function $Z(A)$
 is defined by
 \eqb
  Z(A)
  &=&\int \prod_{x\in\torus}d\bar\psi(x)d\psi(x)
     \exp(-S(\bar\psi,\psi))\nonumber\\
  &=&\det(i\Slash D +MI -\dfrac{r}{2a}W).
 \eqe
 Note that the semi-positivity of $-W$
 \eqb\eqa{uWu}
  (u,-Wu)&=& \sum_{\mu=1}^{d}\norm{(I-T_\mu)u}^2 \ \ge\  0
 \eqe
 implies
 \eqb\eqa{lowerboundH0}
  (u,(i\Slash D+MI-\dfrac{r}{2a}W)u) &\neq& 0\ ,\  u\neq0,
 \eqe
 hence $Z(A)\neq0$.
 \subsection{Lattice Chiral Current}
 We define the lattice chiral current by
 \eqb
  J_\mu(x) &=&
  \dfrac12(\bar\psi(x),\gamma_{d+1}\gamma_\mu(T_\mu\psi)(x))_x
  +\dfrac12((T_\mu\bar\psi)(x),\gamma_{d+1}\gamma_\mu \psi(x))_x\,,
  \nonumber\\
  &&\hspace{50mm} \mu=1,2,\dots,d,
 \eqe
 where the chirality $\gamma_{d+1}$ is by definition
 \eqb\eqa{chirality}
  \gamma_{d+1}&=&i^{d/2}\gamma_1\gamma_2\cdots\gamma_d\,.
 \eqe
 Put
 \eqb\eqa{divchiral}
  Y(x)&=&\dfrac1a\sum_{\mu=1}^{d}(J_\mu(x)-J_\mu(x-ae_\mu))
          -2Mi(\bar\psi(x),\gamma_{d+1}\psi(x))_x,  \ x\in\torus,
 \eqe
 and smear it as
 \eqb
  Y(\xi)&=&a^d\sum_{x\in\torus}\xi(x)Y(x)
 \eqe
 by an arbitrary real-valued smooth function $\xi$
 defined on the continuum torus.
 The functional $Y(x)$ is the `divergence' of lattice chiral current
 with a mass correction.
 Our problem is to calculate $\lim_{a\to0}\langle Y(\xi)\rangle$.
 \subsection{Result}
 Let us state our result.
 \thmb\thma{maintheorem}
  For an arbitrary smooth gauge field $A=(A_\mu)$ and 
  for an arbitrary smooth function $\xi$
  both defined on the continuum torus $\T^d$, 
  it holds that
  \eqb
   \lim_{a\to0}\langle Y(\xi)\rangle
   &=&\quad\dfrac{-2iQ^{d/2}}{(4\pi)^{d/2}(d/2)!}\int_{\T^d}dx\ \xi(x)
      \sum_{\mu_1,\mu_2,\dots,\mu_d=1}^{d}\epsilon_{\mu_1\mu_2\dots\mu_d}
      \nonumber\\
   &&\hspace{40mm}
      F_{\mu_1\mu_2}(x)F_{\mu_3\mu_4}(x)\cdots F_{\mu_{d-1}\mu_d}(x),
  \eqe
  where $\epsilon_{\mu_1\mu_2\dots\mu_d}$ is 
  the totally antisymmetric tensor and
  \eqb\eqa{F}
   F_{\mu\nu}(x)&=&\partial_\mu A_\nu(x)-\partial_\nu A_\mu(x).
  \eqe
 \thme\bparr
 Namely, $\langle Y(x)\rangle$ 
 weakly converges to the Chern class in the continuum limit.
 In the sebsequent sections, we prove this fact with mathematical rigor.
 
 \section{Chiral Oscillation}\seca{chiralosc}
 In this section, we carry out the Grassmann integrations,
 and write the chiral anomaly as a trace of an operator on $\dV$.
 \prpb[\protect\cite{K,SS}]\prpa{H0H1}
 It holds that
 \eqb\eqa{H0H1}
  \langle Y(\xi)\rangle&=&-2i\Tr_{\V^d_a}\left[\Lambda\xi\gamma_{d+1}\right],
 \eqe
 where
 \eqb\eqa{Lambda}
  \Lambda&=&\dfrac12(i\Slash D+MI-\dfrac{r}{2a}W)^{-1}(i\Slash D+MI)+
            \dfrac12(i\Slash D+MI)(i\Slash D+MI-\dfrac{r}{2a}W)^{-1}.
 \eqe
 \prpe
 \prfb
 It is easy to see that
 \eqb
  Y(\xi)&=&a^d\sum_{x\in\torus}\xi(x)Y(x) \\
        &=&(\bar\psi,\xi\gamma_{d+1}\Slash D\psi)
           +(\Slash D\bar\psi,\xi\gamma_{d+1}\psi)
           -2Mi(\bar\psi,\xi\gamma_{d+1}\psi),  \eqa{smearedcurrent}
 \eqe
 where in \equ{smearedcurrent} we used the same symbol $\xi$  
 for the multiplication operator 
 given by $(\xi u)(x)=\xi(x)u(x), x\in\torus$.
 Using \equ{smearedcurrent} and manipulating the Grassmann integrations,
 we obtain \equ{H0H1}.
 Here, \equ{lowerboundH0} ensures 
 the existence of the inverse of $i\Slash D+MI-\frac{r}{2a}W$.
 \prfe
 Taking into account the behavior of the Wilson term $-\frac{r}{2a}W$
 for doublers,
 one may intuitively see that the $\Lambda$ works as a `projection'
 to decouple doublers, and
 according to this picture, one can write the scenario 
 that when the doublers go out of the continuum world,
 they take away a part of lattice chiral current
 and produces the chiral unbalance in the continuum world.
 One may say that this is the origin of chiral anomaly 
 and nonzero index of the Dirac operator.

 Put
 \eqb
   L &=& MI - \dfrac{r}{2a}W,                    \eqa{L}\\
   X_{\mu\nu} &=& D_\mu L^{-1}D_\nu-D_\nu L^{-1}D_\mu\ ,\ 
                  \mu,\nu=1,2,\dots,d,           \eqa{X}\\
   B &=& \dfrac12 \sum_{\mu,\nu=1}^{d}
             \gamma_{\mu}\gamma_{\nu}X_{\mu\nu}, \eqa{B}\\
   C &=& L - \sum_{\mu=1}^{d} D_\mu L^{-1}D_\mu. \eqa{C}
 \eqe
 \prpb[\protect\cite{K,SS}]\prpa{firsttrace}
  It holds that
  \eqb\eqa{firsttrace}
    \langle Y(\xi)\rangle &=& 
      -i\Tr_{\dV}
      \left[\dfrac{r}{2a}(\xi W+W\xi)(C+B)^{-1}\gamma_{d+1}\right].
  \eqe
 \prpe
 \prfb
 Put
  \eqb
    H_0&=&i\Slash D+MI, \\
    H_1&=&i\Slash D+L,
  \eqe
  and rewrite \equ{H0H1} as
  \eqb
    \langle Y(\xi)\rangle &=& 
      -i\Tr_{\V_a^d}\left[\gamma_{d+1}(H_0H_1^{-1}+H_1^{-1}H_0)\xi\right].
  \eqe
  Since $H_0=H_1+\frac{r}{2a}W$ and
  $\Tr_{\dV}\left[\xi\gamma_{d+1}\right] = 0$, it holds that
  \eqb
    \langle Y(\xi)\rangle &=& -i\Tr_{\V_a^d}
      \left[\gamma_{d+1}\frac{r}{2a}(WH_1^{-1}+H_1^{-1}W)\xi\right] .
  \eqe
  Note that
  \eqb
   H_1L^{-1}H_1^* &=& L + \Slash DL^{-1}\Slash D \nonumber\\
                  &=& C + B  
  \eqe
  implies
  \eqb
   H_1^{-1} &=& L^{-1}H_1^*(C+B)^{-1}.
  \eqe
  Furthermore,
  \eqb
    H_1\gamma_{d+1} &=& \gamma_{d+1}H_1^*
  \eqe
  holds,
  while $L$, $\xi W+W\xi$, and $(C+B)^{-1}$ commute with $\gamma_{d+1}$.
  Then we have
  \eqb
    \langle Y(\xi)\rangle &=& 
    -i\Tr_{\dV} \left[\dfrac{r}{2a}
    (\xi W+W\xi)L^{-1}H_1^*(C+B)^{-1}\gamma_{d+1}\right] \nonumber\\
    &=& -i\Tr_{\dV} \left[\dfrac{r}{2a}
    (\xi W+W\xi)L^{-1}\dfrac{H_1+H_1^*}{2}(C+B)^{-1}\gamma_{d+1}\right]
     \nonumber\\
    &=& -i\Tr_{\dV} \left[\dfrac{r}{2a}
    (\xi W+W\xi)(C+B)^{-1}\gamma_{d+1}\right].
  \eqe
 \prfe
 
 \section{Quasi Diagonal Operators}\seca{ndo}
 All the lattice operators appearing in this paper are 
 `quasi diagonal' with respect to the planewave basis.
 In this section, we define the class of quasi diagonal operators,
 give a few examples (\lemu{freeparts}, \lemu{smoothfunction}),
 and prove basic properties (\lemu{sumproduct}, \lemu{inverseoperation}).
 \subsection{Definition of quasi diagonal operators}
 We first fix an orthonormal basis of $\dV$.
 Put
 \eqb 
  \torus^*
    &=& \{p=(p_\mu)\in\real^d\ |\ p_\mu=\frac{2k_\mu+1}{L}\pi, 
                  k_\mu\in\integer,|p_\mu|<\dfrac{2}{a}\pi\}.
 \eqe
 We regard $\torus^*$ as a discrete torus 
 with period $\frac{2}{a}\pi$ and spacing $\frac{2\pi}{L}$
 and denote by $\dist{p}$ the periodic distance on $\torus^*$:
 \eqb
  \dist{p}&=& \max_{\mu=1,2,\dots,d}\dist{p_\mu} \nonumber\\
  &=& \max_{\mu=1,2,\dots,d}\min_{n\in\integer}|p_\mu-\dfrac{2n}{a}\pi|.
 \eqe

 For $p \in \torus^*$,
 define the anti-periodic planewave $u_p$ on $\torus$ with a momentum $p$ by
 \eqb\eqa{planewave}
   u_p(x) &=& L^{-d/2}\exp(ipx) \ , \quad x\in\torus\,.
 \eqe
 Furthermore, 
 let $\chi_\alpha, \alpha=1,2,\dots,2^{d/2},$ be 
 the canonical orthonormal basis of $\complex^{2^{d/2}}$
 with respect to the standard inner product.
 Then, 
 \eqb\eqa{planewavebasis}
  u_{\alpha p} &=& \chi_\alpha\otimes u_p  \in \dV
    \ , \quad \alpha=1,2,\dots,2^{d/2}, p\in\torus^*,
 \eqe
 constitute an orthonormal basis of $\dV$\,:
 \eqb
  (u_{\alpha p},u_{\beta q}) &=& \delta_{\alpha\beta}\delta_{pq}\ ,\ 
  \alpha,\beta=1,2,\dots,2^{d/2},\ p,q\in\torus^*,
 \eqe
 where the inner product $(\ ,\ )$ is the one defined by \equ{innerproduct}.

 Let us formulate the class of quasi diagonal operators.
 Operators such as $T_\mu$ or $C$ acting on $\dV$
 have dependences on the lattice spacing $a$,
 so that when we define such operators,
 we in fact define families of operators
 $T_\mu=T_\mu^{(a)}, a>0,$ or $C=C^{(a)}, a>0$.
 \bparr
 {\bf Definition.}\ 
  Let $k=k(a,p), p\in\torus^*,$ be a non-negative function satisfying
  \eqb\eqa{nd2}
   k(a,p) &\le& c(1+\dist{p-q})^\tau k(a,q)\ ,\ p,q\in\torus^*,
  \eqe
  for some constants $c$ and $\tau$ independent of $a,p,$ and $q$.
  Then, for a family of linear operators $K^{(a)}$ on $\dV$,
  we say that $K^{(a)}$ is 
  {\it a quasi diagonal operator of the order \/} $k$
  and write $K^{(a)}=\O(k)$, if
  \eqb\eqa{nd1}
   |(u_{\alpha p},K^{(a)}u_{\beta q})|
    &\le& c_\sigma \dfrac{k(a,p)}{(1+\dist{p-q})^\sigma}\ ,
  \eqe
  holds for all $\sigma\ge0$,
  $p,q\in\torus^*$, and $\alpha, \beta=1,2,\dots,2^{d/2}$.
  Here the constant $c_\sigma$ may depend on $\sigma$
  but not on $a,p,q,\alpha, \beta$.
 \bparr
 Note that \equ{nd2} and \equ{nd1} imply
 \eqb  
   |(u_{\alpha p},K^{(a)}u_{\beta q})|
     &\le& c'_\sigma \dfrac{k(a,q)}{(1+\dist{p-q})^\sigma} \eqa{nd1'}
 \eqe
 for another constant $c'_\sigma$.
 In other words,
 \eqb
  K^{(a)}=\O(k) &\Longrightarrow& {K^{(a)}}^*=\O(k).
 \eqe
 
 In the following, we often suppress writing $a$-dependences of operators
 explicitly and write $K$ instead of $K^{(a)}$.
 Then we will simply say that $K$ is quasi diagonal and 
 write $K=\O(k)$.
 \subsection{Free parts and multiplication operators}
 Define the `free parts' of $D_\mu, W, L,$ and $C$ by
 \eqb
  D_{\mu0}&=&\dfrac{1}{2a}(T_{\mu0}-T_{\mu0}^*),\\
  W_0&=&-\sum_{\mu=1}^{d}(2-T_{\mu0}-T_{\mu0}^*),\\
  L_0&=&MI-\dfrac{r}{2a}W_0,\\
  C_0&=&L_0-\sum_{\mu=1}^{d}L_0^{-1}D_{\mu0}^2,
 \eqe
 respectively.
 A free part of an operator is equal to the operator
 when the gauge potentials $A_\mu, \mu=1,2,\dots,d,$ vanish.
 The free part of $B$ vanishes.
 Since $-W_0$ is semi-positive (see \equ{uWu}),
 $L_0$ and $C_0$ are positive definite and therefore invertible.

 These operators are {\it exactly diagonal}
 with respect to the planewave basis
 and are the first examples of quasi diagonal operators.
 \bparr
 \lemb\lema{freeparts}
  We have the following.
  \eqb
   T_{\mu0} &=& \O(1),\eqa{ndT0}\\
   T_{\mu0}-I &=& \O(a(1+\dist{p})),\eqa{ndT0-I}\\
   D_{\mu0} &=& \O(1+\dist{p}),\eqa{ndD0}\\
   W_0 &=& \O(a^2(1+\dist{p}^2)),\eqa{ndW0}\\
   L_0 &=& \O(1+a\dist{p}^2),\eqa{ndL0}\\
   L_0^{-1} &=& \O(\dfrac{1}{1+a\dist{p}^2}),\eqa{ndL0inv}\\
   C_0 &=& \O(\dfrac{1+\dist{p}^2}{1+a\dist{p}^2}),\eqa{ndC0}\\
   C_0^{-1} &=& \O(\dfrac{1+a\dist{p}^2}{1+\dist{p}^2})\eqa{ndC0inv}.
  \eqe
 \leme\bparr
 \prfb
  \equ{ndT0}-\equ{ndW0} follow from
  \eqb
   T_{\mu0}u_{\alpha p} &=& e^{iap_\mu}u_{\alpha p}, \\
   D_{\mu0}u_{\alpha p} &=& \dfrac{i}{a}\sin(ap_\mu)u_{\alpha p}, \\
   W_{\mu0}u_{\alpha p} &=& -2\sum_{\mu=1}^{d}
                            (1-\cos(ap_\mu))u_{\alpha p}. \eqa{Wp}
  \eqe
  For example, if $K=T_{\mu0}-I$, 
  then \equ{nd1} is satisfied by putting $k(a,p)=a\dist{p}$.
  We add extra `1' in \equ{ndT0-I} in order to ensure \equ{nd2}:
  the function $k(a,p)=a(1+\dist{p})$ satisfies \equ{nd2},
  because 
  \eqb
   1+\dist{q+r} &\le& 1+\dist{q}+\dist{r} \nonumber\\
     &\le& (1+\dist{r})(1+\dist{q}), \  q,r\in\torus^*.
  \eqe
  For the case $K=W_0$, we need 
  \eqb
    1+\dist{q+r}^2 &\le& 1+2(\dist{q}^2+\dist{r}^2) \nonumber\\
     &\le& 2(1+\dist{r}^2)(1+\dist{q}^2), \  q,r\in\torus^*.
  \eqe
  The other bounds are shown by using
  \eqb
   L_0u_{\alpha p} &=& 
    \left(M+\dfrac{r}{2a}\sum_{\mu=1}^d(1-\cos ap_{\mu})\right)
    u_{\alpha p},  \\
   C_0u_{\alpha p} &=&
      \left(M+\dfrac{r}{2a}\sum_{\mu=1}^d(1-\cos ap_{\mu})
      +\dfrac{\displaystyle\sum_{\mu=1}^d \dfrac{1}{a^2}\sin^2 ap_{\mu}}
       {M+\dfrac{r}{2a}\sum_{\mu=1}^d(1-\cos ap_{\mu})}\right)u_{\alpha p}.
  \eqe
 \prfe\bparr
 A multiplication operator determined by a smooth function on $\T^d$ 
 is another example of quasi diagonal operators.
 \lemb\lema{smoothfunction}
   Let $\phi$ be a multiplication operator defined by
   \eqb
    \phi u(x) &=& \varphi(x)u(x)\ ,\ x\in\torus, u\in\dV,
   \eqe
   for a complex valued smooth (periodic) function $\varphi(x)$
   on the continuum torus $\T^d$.
   Then, we have $\phi=\O(1)$.
   \par
   Furthermore, $\phi=\O(1)$ holds, if
   $\phi$ is defined by
   \eqb
    \phi u(x) &=& \varphi_a(x)u(x)\ ,\ x\in\torus, u\in\dV,
   \eqa{latter1}\eqe
   for a family of functions $\varphi_a(x)$ on $\T^d$ 
   that are `uniformly smooth', i.e.
   \eqb
    \sup_{x\in\T^d}|\partial^\alpha\varphi_a(x)|&<&c^{(\alpha)}\ ,\ 
    \alpha=(\alpha_1,\dots,\alpha_d)\in\{0,1,2,\dots\}^d,
   \eqa{latter2}\eqe
   where $\{c^{(\alpha)}\mid \alpha\in\{0,1,\cdots\}^d\}$
 is a family of constants independent of $a$,
   and $\partial^\alpha$ stands for $\prod_{k=1}^d\partial^{\alpha_k}_k$.
 \leme\bparr
 \prfb
  We show the first part of the lemma.
  Since $\varphi(x)$ is smooth and satisfies
  \eqb
   (u_{\alpha p}, \phi u_{\beta q})
   &=& \delta_{\alpha\beta}a^d\sum_{x\in\torus}e^{i(q-p)x}\varphi(x),\ 
   \alpha,\beta=1,2,\dots,2^{d/2}, p,q\in\torus^*,
  \eqe
  there exists $c_\sigma$ such that
  \eqb
   |(u_{\alpha p},\phi u_{\beta q})|
    &\le& \dfrac{c_\sigma }{(1+\dist{p-q})^\sigma}\ , \sigma\ge0
  \eqe
  for $p,q\in\torus^*$ and for $\alpha,\beta=1,2,\dots,2^{d/2}$.
  Then, $K_a=\phi$ and $k(a,p)=1$ satisfy \equ{nd2} and \equ{nd1}.
 The other part of the lemma can be proved in a similar way.
 \prfe
%
 \subsection{Basic properties}
 The class of quasi diagonal operators
 is closed with respect to summation, product, and the inverse operation
 in the following sense.
 \bparr
 \lemb\lema{sumproduct}
  Let $K_j=\O(k_j), j=1,2$.
  Then, we have $K_1+K_2=\O(k_1+k_2)$ and $K_1K_2=\O(k_1k_2)$.
 \leme\bparr
 The proof is given in \secu{sumproduct}.
 Note that, when we say $K=\O(k)$, it is implied that 
 $k$ has the property \equ{nd2} with some constants $c,\tau$.
 \bparr
 \lemb\lema{inverseoperation}
  Suppose that $K=K_0+K_1$ satisfies the following,
  where `$\const$' stands for some positive constant
  independent of $a$ and $p$:
  \begin{enumerate}
   \item 
         There exists  $k_0$ satisfying
         \eqb\eqa{k0}
           k_0(p) &\le& \const (1+\dist{p})^{\nu}\ ,\  p\in\torus^*,
         \eqe
         for some constant $\nu$ independent of $a$ and $p$, such that
         $K_0$ is exactly of the order $k_0$ in the sense that
         $K_0=\O(k_0)$ and $K_0^{-1}=\O(1/k_0)$ hold.
   \item $K_1$ is of a lower order than $K_0$ in the sense that
         $K_1=\O(k_1)$ and 
         \eqb\eqa{k1/k0}
          \dfrac{k_1(p)}{k_0(p)} &\le& \const(1+\dist{p})^{-\delta}\ ,\ 
            p\in\torus^*,
         \eqe
         for some constant $\delta>0$ independent of $a$ and $p$.
   \item $K^{-1}$ exists and is uniformly bounded, i.e.,
         \eqb\eqa{boundonKinv}
            \norm{K^{-1}} &=& \mathop{\sup_{u\in\dV}}_{u\neq0}
            \dfrac{\norm{K^{-1}u}}{\norm{u}} \le c
         \eqe
         for some constant $c$ independent of $a$.
  \end{enumerate}
  Then we have $K^{-1}=\O(1/k_0)$.
 \leme\bparr
 The proof is given in \secu{inverseoperation}.

 By means of \lemu{sumproduct} and \lemu{inverseoperation},
 we can estimate traces on $\dV$ using the order of operators.
 As may be seen explicitly from  \lemu{freeparts}, such estimates
 provide a mathematical justification of 
 the power counting argument.
 \subsection{Proof of Lemma 4.3}\seca{sumproduct}
 Let us first show $K_1+K_2=\O(k_1+k_2)$.
 Assume
 \eqb
   k_1(a,p) &\le& c_1(1+\dist{p-q})^{\tau_1}k_1(a,q)\ ,\\
   k_2(a,p) &\le& c_2(1+\dist{p-q})^{\tau_2}k_2(a,q)\ ,
    \eqa{k2}\\
   |(u_{\alpha p},K_1u_{\beta q})|
     &\le& c_{1\sigma}\dfrac{k_1(a,p)}{(1+\dist{p-q})^\sigma}\ ,\eqa{K1}\\
   |(u_{\alpha p},K_2u_{\beta q})|
     &\le& c_{2\sigma}\dfrac{k_2(a,p)}{(1+\dist{p-q})^\sigma}\ ,\eqa{K2}
 \eqe
 for $\sigma\ge0,\ p,q\in\torus^*$, and 
 for $\alpha,\beta=1,2,\dots,2^{d/2}$.
 Then, putting 
  $K=K_1+K_2, k=k_1+k_2, \tau=\max(\tau_1,\tau_2), c=\max(c_1,c_2),$
  and $c_\sigma=\max(c_{1\sigma},c_{2\sigma})$,
 we have \equ{nd2} and \equ{nd1}, and consequently, $K_1+K_2=\O(k_1+k_2)$.

 Let us show $K=K_1K_2=\O(k_1k_2)$.
 Put $k=k_1k_2$.
 Then, for $c=c_1c_2$ and for $\tau=\tau_1+\tau_2$, \equ{nd2} holds.
 In order to show \equ{nd1}, we need the following lemma.
 \bparr
 \lemb\lema{convolution}
   If $0\le\rho\le\min\{\sigma,\tau\}$ and $\sigma+\tau-\rho\ge d+1$ hold, 
   we have
   \eqb\eqa{conv}
    \sum_{r\in\torus^*}(1+\dist{p-r})^{-\sigma}(1+\dist{r-q})^{-\tau}
    \le \const \,(1+\dist{p-q})^{-\rho},
   \eqe
   where `$\const$' is independent of $p,q,$ and $a$.
 \leme\bparr
 \prfb
  Decompose the summation in the left hand side of \equ{conv} as
  \eqb
   &&\sum_{r\in\torus^*}(1+\dist{p-r})^{-\sigma}(1+\dist{r-q})^{-\tau}
      \nonumber\\
   &=&\left(\mathop{\sum_{r\in\torus^*}}_{\dist{p-r}\ge\dist{r-q}}
           +\mathop{\sum_{r\in\torus^*}}_{\dist{p-r}<\dist{r-q}}\right)
            (1+\dist{p-r})^{-\sigma}(1+\dist{r-q})^{-\tau}.
   \eqa{convdecomp}
  \eqe
  Let us estimate the first sum.
  If $\dist{p-r}\ge\dist{r-q}$, then it holds that
  \eqb
   && 2\dist{p-r} \ge \dist{p-r}+\dist{r-q} \ge \dist{p-q},
  \eqe
  hence
  \eqb
   &&\mathop{\sum_{r\in\torus^*}}_{\dist{p-r}\ge\dist{r-q}}
            (1+\dist{p-r})^{-\sigma}(1+\dist{r-q})^{-\tau} \nonumber\\
   &\le& \mathop{\sum_{r\in\torus^*}}_{\dist{p-r}\ge\dist{r-q}}
            (1+\dist{p-r})^{-\rho}(1+\dist{p-r})^{-\sigma+\rho}
            (1+\dist{r-q})^{-\tau} \nonumber\\
   &\le& \sum_{r\in\torus^*}
            (1+\frac12\dist{p-q})^{-\rho}(1+\dist{r-q})^{-\tau-\sigma+\rho}
            \nonumber\\
   &\le& \const\,(1+\frac12\dist{p-q})^{-\rho}\,
  \eqe
  where we also used
 the assumptions $\/0\le\rho\le\sigma$ and $\tau+\sigma-\rho\ge d+1$.
  The second sum in the right hand side of \equ{convdecomp} can be
  estimated in a similar way.
 \prfe\bparr
 Using \equ{K1},\equ{K2},\equ{k2} and \equ{conv},
 we have
 \eqb
   |(u_{\alpha p},K_1K_2u_{\beta q})|
     &\le& \sum_{\gamma=1}^{2^{d/2}}\sum_{r\in\torus^*}
      |(u_{\alpha p},K_1u_{\gamma r})|
      |(u_{\gamma r},K_2u_{\beta q})| \nonumber\\
     &\le& \sum_{\gamma=1}^{2^{d/2}}\sum_{r\in\torus^*}
      \dfrac{c_{1\sigma}c_{2\tau}k_1(a,p)k_2(a,r)}
      {(1+\dist{p-r})^\sigma(1+\dist{r-q})^\tau} \nonumber\\
     &\le& \sum_{\gamma=1}^{2^{d/2}}\sum_{r\in\torus^*}
      \dfrac{c_{1\sigma}c_{2\tau}c_2k_1(a,p)k_2(a,p)}
      {(1+\dist{p-r})^{\sigma-\tau_2}(1+\dist{r-q})^\tau} \nonumber\\
     &\le& 
      \const\dfrac{c_{1\sigma}c_{2\tau}c_2k_1(a,p)k_2(a,p)}
      {(1+\dist{p-q})^{\rho}}
 \eqe
 by choosing sufficiently large $\sigma$ and $\tau$ for each $\rho$.
 This implies \equ{nd1} for $K=K_1+K_2$, hence a proof of 
 \lemu{sumproduct} is completed.
 \subsection{Preparation for a Proof of Lemma 4.4}\seca{preparation}
 In this section, we prepare \lemu{JJn}, \lemu{p-qcom}, and \lemu{RKinv}
 for the proof of \lemu{inverseoperation}.

 The idea of the proof of \lemu{inverseoperation} is 
 to estimate the right hand side of the identity
 \eqb
   K^{-1}&=&\sum_{j=0}^{n-1}J^jK_0^{-1}+J^nK^{-1},\   
   n=1,2,\dots, \eqa{iterate1}
 \eqe
 where
 \eqb
  J &=&-K_0^{-1}K_1 .
 \eqe
 \lemu{sumproduct} implies that 
 the first term in the right hand side of \equ{iterate1} is quasi diagonal.
 For the second term, as we shall see later, 
 $J^nK^{-1}$ for a sufficiently large $n$ has a good power counting property
 in spite of the poor information \equ{boundonKinv} on $K^{-1}$.
 The key point is that the order of $J$ is strictly lower than 1:
 \bparr
 \lemb\lema{JJn}
 It holds that
 \eqb
   J &=& \O((1+\dist{p})^{-\delta})\,. \eqa{J}
 \eqe
 Furthermore, for $n\ge\nu/\delta$,
 \eqb
   J^n &=& \O(1/k_0)\ \eqa{Jn}
 \eqe
 holds.
 \leme
 \prfb
 The assumtion \equ{k1/k0} and \lemu{sumproduct} imply \equ{J}.
 Then we have
  \eqb
   J^n &=& \O((1+\dist{p})^{-n\delta})\ ,\ n=1,2,\dots.
  \eqe
  In view of \equ{k0}, 
  we can bound the order of $J^n$ as follows:
  \eqb
   (1+\dist{p})^{-n\delta} &\le& \const
     \dfrac{(1+\dist{p})^{\nu-n\delta}}{k_0(p)}.
  \eqe
  Choosing $n\ge\nu/\delta$,
  we obtain \equ{Jn}.
 \prfe\bparr
 The above lemma implies that 
 multiplying by $J$ improves the power counting properties of operators.

 In order to determine the order of the second term 
 in the right hand side of \equ{iterate1},
 we have to bound
 \eqb
   \dist{p_\mu-q_\mu}^\rho|(u_{\alpha p},J^nK^{-1}u_{\beta q})|
 \eqe
 for $\rho\ge0, p,q\in\torus^*$, and for $\alpha,\beta=1,2,\dots,2^{d/2}$.
 To this end, we use commutators with $\frac1a T_{\mu0}$.
 Let us write
 \eqb
  \ad_\mu(R) &=& \left[\frac1a T_{\mu0}, R\right]
 \eqe
 for an operator $R$ on $\dV$.
 \bparr
 \lemb\lema{p-qcom}
   For an operator $R$ on $\dV$, it holds that
   \eqb\eqa{p-q}
    |(u_{\alpha p},\ad_\mu(R)u_{\beta q})|
    &&\left\{ \begin{array}{l} 
     \le \dist{p_\mu-q_\mu}|(u_{\alpha p},Ru_{\beta q})|, \\
     \ge\frac{1}{\pi} \dist{p_\mu-q_\mu}|(u_{\alpha p},Ru_{\beta q})|,
     \end{array} \right. \\ \nonumber
    &&\quad\quad p,q\in\torus^*,\ \alpha,\beta=1,2,\dots,2^{d/2},\ 
     \mu=1,2,\dots,d.
   \eqe
   Furthermore, if $R=\O(r)$, then $\ad_\mu(R)=\O(r)$.
 \leme
 \prfb
  The bounds \equ{p-q} follows from
  \eqb
   |(u_{\alpha p},\ad_\mu(R)u_{\beta q})|
   &=& \dfrac{1}{a}|(\exp(ip_\mu a)-\exp(iq_\mu a))|
       |(u_{\alpha p},Ru_{\beta q})| \nonumber\\
   &=& \dfrac{2}{a}|\sin(\frac12(p_\mu-q_\mu)a)|
       |(u_{\alpha p},Ru_{\beta q})|.
  \eqe
  The estimate $\ad_\mu(R)=\O(r)$ follows from \equ{p-q} and \equ{nd1}.
 \prfe\bparr
 \bparr
 \lemb\lema{RKinv}
 \begin{enumerate}
  \item For $R=\O(r)$, we have
  \eqb\eqa{bound(p-q)^0}
    |(u_{\alpha p}, RK^{-1}u_{\beta q})| &\le& \const r(p)\ ,\ 
    p,q\in\torus^*, \ \alpha,\beta=1,2,\dots,2^{d/2}, 
  \eqe
  where `$\const$' is a constant independent of $a,p$, and $q$.
 \item Assume that $R=\O(r)$ is at most of the same order as $K$, i.e.,
  \eqb\eqa{r<k0}
   r(p) &\le& k_0(p)\ ,\ p\in\torus^*.
  \eqe
  Then, we have
  \eqb
  \sum_{s'\in\torus^*}|(u_{\alpha s'}, RK^{-1}u_{\beta s})|
  &\le& \const\ ,\ 
  s\in\torus^*, \ \alpha,\beta=1,2,\dots,2^{d/2}, \eqa{sumRKinv}
  \eqe
  where `$\const$' is a constant independent of $a$ and $s$.
 \end{enumerate}
 \leme
 \prfb
  (1)\ \equ{nd1} and \equ{boundonKinv} imply
  \eqb
   |(u_{\alpha p}, Ru_{\gamma s})| &\le& 
     \dfrac{\const r(p)}{(1+\dist{p-s})^{d+1}}\ ,\\
   |(u_{\gamma s},K^{-1}u_{\beta p})| &\le& c,
  \eqe
  for $p,s\in\torus^*$ and for $\alpha,\beta,\gamma=1,2,\dots,2^{d/2}$.
  Then, we have
  \eqb
   |(u_{\alpha p}, RK^{-1}u_{\beta q})| &\le& 
    \sum_{s\in\torus^*}|(u_{\alpha p}, Ru_{\gamma s})|
                       |(u_{\gamma s},K^{-1}u_{\beta q})| \nonumber\\
   &\le& 
   \sum_{s\in\torus^*}\dfrac{\const r(p)}{(1+\dist{p-s})^{d+1}}\nonumber\\
   &\le& \const r(p),
  \eqe
  where we used
  \eqb
   \sum_{s\in\torus^*}(1+\dist{s})^{-d-1} &\le& \const
  \eqe

  \noindent
  (2)\ Using \equ{iterate1}, we have
  \eqb
  &&\sum_{s'\in\torus^*}|(u_{\alpha s'}, RK^{-1}u_{\beta s})| \nonumber\\
  &\le& 
   \sum_{s'\in\torus^*}\sum_{j=0}^{n-1}
    |(u_{\alpha s'}, RJ^jK_0^{-1}u_{\beta s})| 
    + \sum_{s'\in\torus^*}|(u_{\alpha s'}, RJ^nK^{-1}u_{\beta s})| .
  \eqe
  Let us bound the right hand side of the above inequality.
  For the first term, \equ{J} implies
  \eqb\eqa{J1}
   J=\O(1),
  \eqe
  hence
  \eqb\eqa{RJjK0inv}
   RJ^jK_0^{-1} &=& \O(\dfrac{r}{k_0})
  \eqe
  holds.
  Then, from \equ{RJjK0inv}, \equ{nd1'}, and \equ{r<k0}, we obtain
  \eqb
   \sum_{s'\in\torus^*}\sum_{j=0}^{n-1}
    |(u_{\alpha s'}, RJ^jK_0^{-1}u_{\beta s})| 
   &\le& \const\dfrac{r(s)}{k_0(s)} 
    \sum_{s'\in\torus^*}(1+\dist{s'-s})^{-d-1} \nonumber\\
   &\le& \const\ ,\quad s\in\torus^*,
  \eqe
  for a `$\const$' independent of $a$ and $s$.
  For the second term, note that \equ{k0}
 and $r(p) \le k_0(p)$ imply that by choosing a sufficiently large $n$, 
  we can make the order $r(p)(1+\dist{p})^{-n\delta}$ of $RJ^n$
  lower than $(1+\dist{p})^{-d-1}$.
  Then, using \equ{boundonKinv} for $K^{-1}$ and \equ{nd2}
 for $RJ^n$ with $\sigma=d+1$, we have
  \eqb
   \sum_{s'\in\torus^*}|(u_{\alpha s'}, RJ^nK^{-1}u_{\beta s})|
   &\le&
    \sum_{s',s''\in\torus^*}\sum_{\gamma=1}^{2^{d/2}}
    |(u_{\alpha s'}, RJ^nu_{\gamma s''})||(u_{\alpha s''}, K^{-1}u_{\beta s})|
    \nonumber\\
   &\le&\sum_{s',s''\in\torus^*}\sum_{\gamma=1}^{2^{d/2}}
    \const(1+\dist{s'})^{-d-1}(1+\dist{s'-s''})^{-d-1}\nonumber\\
   &\le&\const .
  \eqe
 \prfe\bparr
%
 \subsection{Proof of Lemma 4.4}\seca{inverseoperation}
 In order to show \lemu{inverseoperation},
 it suffices to estimate the right hand side of \equ{iterate1}.
 For the first term, using \equ{J1}, we have the following bound:
 \eqb
  \sum_{j=0}^{n-1}J^jK_0^{-1} &=& \O(1/k_0).
 \eqe

 For the second term, we have to estimate
 \eqb\eqa{(p-q)^rho}
   \dist{p_\mu-q_\mu}^\rho|(u_{\alpha p},J^nK^{-1}u_{\beta q})|
 \eqe
 for $\rho\ge0, p,q\in\torus^*$ and for $\alpha,\beta=1,2,\dots,2^{d/2}$.

 If $\rho=0$,
 \equ{Jn} and \equ{bound(p-q)^0} yield
 \eqb\eqa{Bound(p-q)^0}
   |(u_{\alpha p}, J^nK^{-1}u_{\beta p})|
   &\le& \dfrac{\const}{k_0(p)}\ ,\ 
   p,q\in\torus^*, \ \alpha,\beta=1,2,\dots,2^{d/2}.
 \eqe
 Now assumue $\rho>0$. Then
 \lemu{p-qcom} implies
 \eqb\eqa{p-qtoad}
  \dist{p_\mu-q_\mu}^\rho|(u_{\alpha p},J^nK^{-1}u_{\beta q})|
  &\le&  \pi^\rho|(u_{\alpha p},\ad_\mu^\rho(J^nK^{-1})u_{\beta q})|\ ,
  \nonumber\\
  && \quad p,q\in\torus^*, \ \alpha,\beta=1,2,\dots,2^{d/2}.
 \eqe
 Here we note the equalities
 \eqb
    \ad_\mu^\rho(J^nK^{-1}) &=& 
\sum_{m=0}^{\rho} \left(\begin{array}{c} \rho \\ m \end{array} \right)\,
 \ad_\mu^{\rho-m}(J^n)\ad_\mu^m(K^{-1})
 \eqe
 and
 \eqb\eqa{ad^m}
    &&\ad_\mu^m(K^{-1}) \nonumber\\
    &=&
    \sum_{\ell=1}^{m}\sum_{(m_1,m_2,\dots,m_\ell)}
    \dfrac{(-1)^\ell m!}{m_1!m_2!\cdots m_\ell!}
    K^{-1}\ad_\mu^{m_1}(K)K^{-1}\ad_\mu^{m_2}(K)K^{-1}\cdots
    \ad_\mu^{m_\ell}(K)K^{-1},\nonumber\\
    &&\hspace{70mm}m=1,2,\dots,
 \eqe
 where the summation $\sum_{(m_1,m_2,\dots,m_\ell)}$
 is taken over all $(m_1,m_2,\dots,m_\ell)$'s such that
 \eqb
         m_1+m_2+\cdots+m_\ell &=& m, \\
         m_1,m_2,\dots,m_\ell &\ge& 1.
 \eqe
 Combining them, we obtain
 \eqb\eqa{ad(JnKinv)}
    \ad_\mu^\rho(J^nK^{-1}) &=& \ad_\mu^{\rho}(J^n)K^{-1} \nonumber\\
&& +\sum_{m=1}^{\rho} \left(\begin{array}{c} \rho \\ m \end{array} \right)\,
 \sum_{\ell=1}^{m}\sum_{(m_1,m_2,\dots,m_\ell)}
     \dfrac{(-1)^\ell m!}{m_1!m_2!\cdots m_\ell!}\nonumber\\
     && \quad \ad_\mu^{\rho-m}(J^n)K^{-1}
     \ad_\mu^{m_1}(K)K^{-1}\ad_\mu^{m_2}(K)K^{-1}\cdots
     \ad_\mu^{m_\ell}(K)K^{-1}.\nonumber\\
 \eqe
 Then, it suffices to bound
 $\ad_\mu^{m'}(J^n)K^{-1}$ and $\ad_\mu^{m}(K)K^{-1}$.

 Note that \equ{Jn} and the latter part of \lemu{p-qcom} imply
 $\ad_\mu^{m'}(J^n)=\O(1/k_0)$ for sufficiently large $n$.
 Then, using \equ{bound(p-q)^0}, we have
 \eqb\eqa{adJnKinv}
  |(u_{\alpha p}, \ad_\mu^{m'}(J^n)K^{-1}u_{\beta s})|
   &\le& \dfrac{\const}{k_0(p)}\ ,\ 
   p,s\in\torus^*,\ \alpha,\beta=1,2,\dots,2^{d/2}.
 \eqe
 Furthermore, the assumptions on the orders of $K_0$ and $K_1$ imply
 $K=\O(k_0)$. This fact and \lemu{p-qcom} and \equ{sumRKinv} imply
 \eqb\eqa{adKKinv}
  \sum_{s\in\torus^*}\sum_{\beta=1}^{2^{d/2}}
   |(u_{\beta s}, \ad_\mu^{m}(K)K^{-1}u_{\gamma s'})| &\le& \const\ ,\ 
   s'\in\torus^*,\ \gamma=1,2,\dots,2^{d/2}.
 \eqe
 Then, 
 \equ{p-qtoad}, \equ{ad(JnKinv)}, \equ{adJnKinv}, and \equ{adKKinv} imply
 \eqb
  \dist{p_\mu-q_\mu}^\rho|(u_{\alpha p}, J^nK^{-1} u_{\beta q})|
  &\le& \dfrac{\const}{k_0(p)}\ ,\nonumber\\
  \rho>0,\ p,q&\in&\torus^*,\ \alpha,\beta=1,2,\dots,2^{d/2}.
 \eqe
 This together with \equ{Bound(p-q)^0} yields $J^nK^{-1}=\O(1/k_0)$,
 which concludes the analysis on 
 the second term in the right hand side of \equ{iterate1},
 hence the proof of \lemu{inverseoperation}.
 
 \section{Orders of Operators}\seca{order}
 Using the framework of quasi diagonal operators,
 we determine the orders of lattice operators.
 \subsection{Orders of Interaction parts}
 Interaction parts of the operators 
 $T_\mu, D_\mu, W, L,$ and $C$ are by definition
 \eqb
  T_{\mu1} &=& T_\mu-T_{\mu0}, \\
  D_{\mu1} &=& D_\mu-D_{\mu0}, \\
  W_1 &=& W-W_0, \\
  L_1 &=& L-L_0, \\
  C_1 &=& C-C_0.
 \eqe
 Namely,
 \eqb
  T_{\mu1} &=& (U_\mu-I)T_{\mu0},                  \eqa{Tmu1}\\
  D_{\mu1} &=& \dfrac{1}{2a}(T_{\mu1}-T^*_{\mu1}), \eqa{Dmu1}\\
  W_1 &=& \sum_{\mu=1}^{d}
          \left((U_\mu-I)(T_{\mu0}-I)+(T^*_{\mu0}-I)(U^*_\mu-I)
          +(U_\mu+U^*_\mu-2I)\right),              \eqa{W1}\\
  L_1 &=& -\dfrac{r}{2a}W_1,                       \eqa{L1}\\
  C_1 &=& L_1-\sum_{\mu=1}^d 
             \left(D_{\mu}(L^{-1}-L_0^{-1})D_{\mu} 
             +D_{\mu1}L_0^{-1}D_{\mu}
             +D_{\mu0}L_0^{-1}D_{\mu1} \right) .   \eqa{C1}
 \eqe
 \bparr
 \lemb\lema{perturbativepart}
 We have the following order estimates.
 \begin{enumerate}
  \item The orders of the interaction parts:
  \eqb
   T_{\mu1} &=& \O(a), \\
   D_{\mu1} &=& \O(1), \eqa{D1O}\\
   W_1 &=& \O(a^2(1+\dist{p})), \eqa{W1O}\\
   L_1 &=& \O(a(1+\dist{p})). \eqa{L1O}
  \eqe
  \item The orders of the full operators:
  \eqb
   T_{\mu} &=& \O(1), \\
   D_{\mu} &=& \O(1+\dist{p}), \eqa{DmuO}\\
   W &=& \O(a^2(1+\dist{p}^2)), \eqa{WO}\\
   L &=& \O(1+a\dist{p}^2).
  \eqe
 \end{enumerate}
 \leme
 \prfb
 (1)\ follows from \equ{Tmu1}--\equ{L1} and \lemu{sumproduct}
 together with \lemu{freeparts} and 
 \eqb
  U_\mu-I, U^*_\mu-I &=& \O(a) ,\\
  U_\mu+U^*_\mu-2I &=& \O(a^2).
 \eqe
 (2)\ follows from (1) and \lemu{freeparts}.\parr
 \prfe
 \subsection{Orders of $L^{-1}$ and $C^{-1}$}
%
 \lemb\lema{L}
  The operator $L$ is positive definite and satisfies
  \eqb
   (u,Lu) &\ge& M\norm{u}^2, \eqa{uLu}\\
   \norm{L^{-1}} &\le& M^{-1}. \eqa{normLinv}
  \eqe
  Furthermore, it holds that
  \eqb
    L^{-1} &=& \O(\dfrac{1}{1+a\dist{p}^2}), \eqa{LinvO}\\
    L^{-1}-L_0^{-1} &=& \O(\dfrac{a(1+\dist{p})}{(1+a\dist{p}^2)^2}).
    \eqa{Linv-L0inv}
  \eqe
 \leme
 \prfb
  The semi-positivity of $-W$ (see \equ{uWu}) yields \equ{uLu},
  from which \equ{normLinv} follows.
  Furthermore, we have \equ{LinvO} from
  \equ{ndL0}, \equ{ndL0inv}, \equ{L1O}, and \lemu{inverseoperation},
  because
  \eqb
   \dfrac{a(1+\dist{p})}{1+a\dist{p}^2} &\le& \dfrac{2}{1+\dist{p}}.
  \eqe
  Finally, estimating the right hand side of
  \eqb
    L^{-1}-L_0^{-1} &=& -L^{-1}L_1L_0^{-1},
  \eqe
  we obtain \equ{Linv-L0inv}.
 \prfe\bparr
 \lemb\lema{C}
  The operator $C$ is positive definite and satisfies
  \eqb
   (u,Cu) &\ge& M\norm{u}^2, \eqa{uCu}\\
   \norm{C^{-1}} &\le& M^{-1}. \eqa{normCinv}
  \eqe
  Furthermore, it holds that
  \eqb
    C_1 &=& \O(\dfrac{1+\dist{p}}{1+a\dist{p}^2}), \eqa{C1O}\\
    C &=& \O(\dfrac{1+\dist{p}^2}{1+a\dist{p}^2}), \eqa{CO}\\
    C^{-1} &=& \O(\dfrac{1+a\dist{p}^2}{1+\dist{p}^2}). \eqa{CinvO}
  \eqe
 \leme
 \prfb
  In the right hand side of \equ{C}, 
  $L$ is positive definite (\lemu{L}) and
  \eqb
    D_\mu^* &=& -D_\mu \  ,\  \mu=1,2,\dots,d.
  \eqe
  Then, \equ{uCu} and \equ{normCinv} hold.
  Applying \lemu{freeparts}, \lemu{perturbativepart}, and \lemu{L}
  to operators in the right hand side of \equ{C1}, 
  we have  \equ{C1O}, which, with \equ{ndC0}, implies \equ{CO}.
  Finally, we obtain \equ{CinvO} 
  from \equ{ndC0}, \equ{ndC0inv}, and \equ{normCinv}.
 \prfe
 \subsection{Order of $X_{\mu\nu}$}
 We begin with the following.
 \lemb\lema{Tphi}
 If $\phi$ is a multiplication operator which satisfies
 \equ{latter1} and \equ{latter2},
   we have, for $\mu=1,2,\dots,d$,
   \eqb
    \left[T_{\mu0},\phi\right] &=& \O(a), \eqa{comT0phi}\\
    \left[T^*_{\mu0},\phi\right] &=& \O(a), \eqa{comT0sphi}\\
    \left[T_{\mu0}+T^*_{\mu0},\phi\right] &=& \O(a^2(1+\dist{p})).
                                          \eqa{comT0T0sphi}
   \eqe
 \leme
 \prfb
  We bound the right hand side of
  \eqb
    [T_{\mu0},\phi] &=& (T_{\mu0}\phi T^*_{\mu0}-\phi)T_{\mu0}.
    \eqa{precomT1T0}
  \eqe
  Since $T_{\mu0}\phi T^*_{\mu0}$ is a multiplication operator
  determined by the translation of $\varphi$,
  it holds that
  \eqb
    T_{\mu0}\phi T^*_{\mu0}-\phi &=& \O(a).
  \eqe
  This proves \equ{comT0phi}.
  The estimate \equ{comT0sphi} is obtained in a similar way.
  The estimate \equ{comT0T0sphi} follows from
  \eqb
    \left[T_{\mu0}+T^*_{\mu0},\phi\right]T_{\mu0}
    &=& 2a(T_{\mu0}\phi T^*_{\mu0}-\phi)D_{\mu0}\nonumber\\
    &&\quad
       +(T_{\mu0}\phi T^*_{\mu0}+T^*_{\mu0}\phi T_{\mu0}-2\phi)
         T^*_{\mu0}                                  \eqa{T0T0sT1}
  \eqe
  with a help of
  \eqb
   T_{\mu0}\phi T^*_{\mu0}+T^*_{\mu0}\phi T_{\mu0}-2\phi
    &=& \O(a^2).
  \eqe
 \prfe\bparr
 The refined bound \equ{comT0T0sphi} due to 
 the cancellation between $T_{\mu0}$ and $T^*_{\mu0}$
 is essential in the proof of \prpu{zj}.
 \bparr
 \lemb
 We have the following order estimates:
 \eqb
   \left[D_{\mu1},D_{\nu0}\right] &=& \O(1), \eqa{comD1D0}\\
   \left[D_{\mu1},D_{\nu1}\right] &=& \O(a), \eqa{comD1D1}\\
   \left[D_{\mu},D_{\nu}\right] &=& \O(1), \eqa{comDD}\\
   \left[L_0,D_{\nu1}\right] &=& \O(a(1+\dist{p})), \eqa{comL0D1}\\
   \left[L_1,D_{\nu0}\right] &=& \O(a(1+\dist{p})), \eqa{comL1D0}\\
   \left[L_1,D_{\nu1}\right] &=& \O(a), \eqa{comL1D1}\\
   \left[L,D_{\nu}\right] &=& \O(a(1+\dist{p})), \eqa{comLD}\\
   X_{\mu\nu} &=& \O(\dfrac{a(1+\dist{p}^2)}{(1+a\dist{p}^2)^2}), \eqa{XO}
 \eqe 
 where $\mu,\nu=1,2,\dots,d$.
 \leme
 \prfb
  Put $\eta_\mu=U_\mu-I$.
  Then, $\eta_\mu=\O(a)$ and 
  \eqb\eqa{preprecomT1T0}
   \left[T_{\mu1},T_{\nu0}\right] &=& [\eta_\mu,T_{\nu0}]T_{\mu0}
  \eqe
  hold.
  Using \equ{comT0phi} for $\phi=\frac{1}{a}\eta_\mu$ 
  (the functions $U_\mu(x)$ and $\eta_\mu(x)$ can be extended
   on the whole $\T^d$),
  we obtain
  \eqb
   \left[T_{\mu1},T_{\nu0}\right] &=& \O(a^2).
  \eqe
  Similarly, we have
  \eqb
   \left[T^*_{\mu1},T_{\nu0}\right]\ ,\ 
   \left[T_{\mu1},T^*_{\nu0}\right]\ ,\ 
   \left[T^*_{\mu1},T^*_{\nu0}\right] &=& \O(a^2).
  \eqe
  Then, \equ{comD1D0} holds.
  The estimates
  \eqb
    \left[T_{\mu1},T_{\nu1}\right] \ =\  
     -\eta_\mu[\eta_\nu,T_{\mu0}]T_{\nu0}+\eta_\nu[\eta_\mu,T_{\nu0}]T_{\mu0}
     &=& \O(a^3), \\
    \left[T_{\mu1}^*,T_{\nu1}\right],\ 
    \left[T_{\mu1},T_{\nu1}^*\right],\  
    \left[T_{\mu1}^*,T_{\nu1}^*\right]  &=& \O(a^3),
  \eqe
  are obtained similarly, from which we have \equ{comD1D1}.
  The estimate \equ{comDD} follows from \equ{comD1D0} and \equ{comD1D1},
  because $D_{\mu0}$ and $D_{\nu0}$ commutes.

  Next, using
  \eqb
   \left[T_{\mu0}+T^*_{\mu0},T_{\nu1}\right] 
    &=& \left[T_{\mu0}+T^*_{\mu0},\eta_\nu\right]T_{\nu0}\,, 
                                                 \eqa{preT0T0sT1}\\
   \left[T_{\mu0},T_{\nu1}+T^*_{\nu1}\right] 
    &=& T_{\nu0}\left[T_{\mu0},\eta_\nu+\eta^*_\nu\right]
       -2aD_{\nu0}\left[T_{\mu0},\eta^*_\nu\right],
                                                 \eqa{T0T1T1s}
  \eqe
  and \lemu{Tphi}, and the fact that $\eta_\nu+\eta^*_\nu=\O(a^2)$,
  we have
  \eqb
   \left[T_{\mu0}+T^*_{\mu0},T_{\nu1}\right] &=& \O(a^3(1+\dist{p})),\\
   \left[T_{\mu0},T_{\nu1}+T^*_{\nu1}\right] &=& \O(a^3(1+\dist{p})).
  \eqe
  Similarly,
  \eqb
   \left[T_{\mu0}+T^*_{\mu0},T^*_{\nu1}\right] &=& \O(a^3(1+\dist{p})),\\
   \left[T^*_{\mu0},T_{\nu1}+T^*_{\nu1}\right] &=& \O(a^3(1+\dist{p}))
  \eqe
  hold and we obtain \equ{comL0D1} and \equ{comL1D0}.
  Furthermore, \equ{comL1D1} is shown by estimating the right hand sides of
  \eqb
   \left[T_{\mu1}+T^*_{\mu1},T_{\nu1}\right] &=& 
     \eta_\mu[T_{\mu0},\eta_\nu]T_{\nu0}
     +\eta_\nu[\eta_\mu,T_{\nu0}]T_{\mu0}  \nonumber\\
    &&\quad+[T^*_{\mu0},\eta_\nu]\eta^*_\mu T_{\nu0}
     +\eta_\nu T^*_{\mu0}[\eta^*_\mu,T_{\nu0}]
  \eqe
  and its adjoint.
  The estimate \equ{comLD} follows from \equ{comL0D1}--\equ{comL1D1}.
  Finally, note that
 \eqb\eqa{Xcom}
   X_{\mu\nu}&=& D_\mu L^{-1}[D_\nu,L]L^{-1}
       -D_\nu L^{-1}[D_\mu,L]L^{-1}
       +[D_\mu,D_\nu]L^{-1}\,.
 \eqe
 Estimating the right hand side of \equ{Xcom}
  by means of \equ{DmuO}, \equ{LinvO}, \equ{comDD}, and \equ{comLD},
  we obtain \equ{XO}.
 \prfe\bparr
 \lemb
  The operator $C+ B$ is positive definite and satisfies
  \eqb
   (u,(C+ B)u) &\ge& M\norm{u}^2, \eqa{uC+Bu}\\
   \norm{(C+ B)^{-1}} &\le& M^{-1}. \eqa{normC+Binv}
  \eqe
  Furthermore, it holds that
  \eqb
    B &=& \O(\dfrac{a(1+\dist{p}^2)}{(1+a\dist{p}^2)^2}), \eqa{BO}\\
    BC^{-1} &=& \O(\dfrac{a}{1+a\dist{p}^2}) ,            \eqa{BCinvO}\\
    (C+ B)^{-1} &=& \O(\dfrac{1+a\dist{p}^2}{1+\dist{p}^2}). 
                                                          \eqa{B+CinvO}
  \eqe
 \leme
 \prfb
  \equ{uC+Bu} and \equ{normC+Binv} are obtained from
  \eqb
   (u,(C+ B)u) &=& 
   (u,Lu)+\sum_{\mu=1}^{d}(D_\mu u,L^{-1}D_\mu u)
  \eqe
  and \equ{uLu}.
  \equ{BO} is a consequence of \equ{B} and \equ{XO}.
  \equ{BCinvO} follows from \equ{BO} and \equ{CinvO}.
  \lemu{inverseoperation} then yields \equ{B+CinvO}.
 \prfe
 \subsection{Leading Term of $X_{\mu\nu}$}
 We extract the leading term of $X_{\mu\nu}$ 
 and bound the remainder.
 Put
 \eqb\eqa{Emu0}
  E_{\mu0} &=& \dfrac12(T_{\mu0}+T^*_{\mu0})\ ,\ \mu=1,2,\dots,d.
 \eqe
 \lemb
  For $\mu,\nu=1,2,\dots,d$, we have
  \eqb
   \left[D_\mu,D_\nu\right]+iQF_{\mu\nu}E_{\mu0} E_{\nu0} &=& \O(a),
     \eqa{comDDlead}\\
   \left[L,D_\mu\right]+iarQ\sum_{\rho=1}^d F_{\mu\rho}D_{\rho0}E_{\mu0} 
    &=& \O(a),
     \eqa{comLDlead}
  \eqe
  where $Q$ is the charge and 
  $F_{\mu\nu}$ is the field strength defined by \equ{F}.
 \leme
 \prfb
  Put $\eta_\mu=U_\mu-I$.
  Recalling \equ{U}, we have
  \eqb
   \eta_\mu-T_{\nu0}\eta_\mu T^*_{\nu0}
    &=& -a\partial_\nu\eta_\mu+\O(a^3) \nonumber\\
    &=& -ia^2Q\partial_\nu A_\mu+\O(a^3).
  \eqe
  Then, \equ{preprecomT1T0} and \equ{precomT1T0} yield
  \eqb
   \left[T_{\mu1},T_{\nu0}\right]
    &=& ia^2Q(\partial_\nu A_\mu)T_{\mu0}T_{\nu0}+\O(a^3).
  \eqe
  Similarly, we have
  \eqb
   \left[T_{\mu1},T^*_{\nu0}\right]
    &=& -ia^2Q(\partial_\nu A_\mu)T_{\mu0}T^*_{\nu0}+\O(a^3),\\
   \left[T^*_{\mu1},T_{\nu0}\right]
    &=& -ia^2Q(\partial_\nu A_\mu)T^*_{\mu0}T_{\nu0}+\O(a^3),\\
   \left[T^*_{\mu1},T^*_{\nu0}\right]
    &=& ia^2Q(\partial_\nu A_\mu)T^*_{\mu0}T^*_{\nu0}+\O(a^3),
  \eqe
  hence
  \eqb
   \left[D_{\mu1},D_{\nu0}\right]
    &=& iQ(\partial_\nu A_\mu)E_{\mu0}E_{\nu0}+\O(a).
  \eqe
  This together with \equ{comD1D1} yields \equ{comDDlead}.
  Let us show \equ{comLDlead}.
  \equ{T0T0sT1} and \equ{preT0T0sT1} yield
  \eqb
   \left[T_{\rho0}+T^*_{\rho0},T_{\mu1}\right] 
    &=& -2ia^3Q(\partial_\rho A_\mu)D_{\rho0}T_{\mu0}+\O(a^3).
  \eqe
  Similarly, we have
  \eqb
   \left[T_{\rho0}+T^*_{\rho0},T^*_{\mu1}\right] 
    &=& 2ia^3Q(\partial_\rho A_\mu)D_{\rho0}T^*_{\mu0}+\O(a^3).
  \eqe
  Then,
  \eqb\eqa{comL0D1lead}
   \left[L_0,D_{\mu1}\right] 
    &=& iarQ\sum_{\rho=1}^d(\partial_\rho A_\mu)D_{\rho0}E_{\mu0}+\O(a)
  \eqe
  follows.
  Furthermore, \equ{T0T1T1s} implies
  \eqb\eqa{comL1D0lead}
   \left[L_1,D_{\mu0}\right] 
    &=& -iarQ\sum_{\rho=1}^d(\partial_\mu A_\rho)D_{\rho0}E_{\mu0}+\O(a).
  \eqe
  Then, \equ{comLDlead} follows from
  \equ{comL0D1lead}, \equ{comL1D0lead}, and \equ{comL1D1}.
 \prfe\bparr
 \lemb
  It holds that
  \eqb
   X_{\mu\nu} 
   &+&iQ\left(F_{\mu\nu}E_{\mu0} E_{\nu0} L_0^{-1}
          +ar\sum_{\rho=1}^dF_{\rho\nu}L_0^{-2}D_{\rho0}D_{\mu0}E_{\nu0}
          -ar\sum_{\rho=1}^dF_{\rho\mu}L_0^{-2}D_{\rho0}D_{\nu0}E_{\mu0}
   \right)\nonumber\\
   &=& \O(\dfrac{a(1+\dist{p})}{(1+a\dist{p}^2)^2})  \eqa{leadX}
  \eqe
  for $\mu,\nu=1,2,\dots,d$.
 \leme
 \prfb
  We extract the leading term from the right hand side of \equ{Xcom}.
  Firstly, \equ{comDDlead} and \equ{Linv-L0inv} imply
  \eqb
   \left[D_\mu,D_\nu\right]L^{-1}&=&-iQF_{\mu\nu}E_{\mu0}E_{\nu0}L_0^{-1}
    +\O(\dfrac{a(1+\dist{p})}{(1+a\dist{p}^2)^2}).
  \eqe
  Similarly, from \equ{comLDlead}, we have
  \eqb\eqa{DLDLLinv}
   D_{\mu}L^{-1}\left[D_\nu,L\right]L^{-1}
   &=&-iarQ\sum_{\rho}D_{\mu0}L_0^{-1}F_{\rho\nu}D_{\rho0}E_{\nu0}L_0^{-1}
    +\O(\dfrac{a(1+\dist{p})}{(1+a\dist{p}^2)^2}).
  \eqe
  In the right hand side of the above equality,
  we may move $F_{\rho\nu}$ to the front of the term
  by using the following lemma.
  (The difference is absorbed 
  in the second term of the right hand side of \equ{DLDLLinv}.)
  Thus we have the desired result.
 \prfe\bparr
 \lemb\lema{comfreefunc}
 If $\phi$ is a multiplication operator which satisfies
 \equ{latter1} and \equ{latter2},
   we have, for $\mu=1,2,\dots,d$,
   \eqb
    \left[D_{\mu0},\phi\right] &=& \O(1), \eqa{comDmu0phi}\\
    \left[E_{\mu0},\phi\right] &=& \O(a), \eqa{comEmu0phi}\\
    \left[W_{0},\phi\right] &=& \O(a^2(1+\dist{p})), \eqa{comW0phi}\\
    \left[L_{0}^{-1},\phi\right] &=& 
     \O(\dfrac{1}{(1+a\dist{p}^2)(1+\dist{p})}), \eqa{comL0invphi}\\
    \left[C_{0}^{-1},\phi\right] 
     &=& \O(\dfrac{1+a\dist{p}^2}{(1+\dist{p})^3}). \eqa{comC0invphi}
   \eqe
 \leme
 \prfb
  The estimates \equ{comDmu0phi} -- \equ{comW0phi}
  follow from \equ{comT0phi} -- \equ{comT0T0sphi}.
  The commutators with $L_{0}^{-1}$ and with $C_{0}^{-1}$ are estimated
  using commutators with $L_{0}$ and with $C_{0}$, respectively.
 \prfe
 
 \section{Proof of \protect\thmu{maintheorem}}\seca{prooftheorem}
 \subsection{Irrelevant Terms}
 If we formally expand $(C+B)^{-1}$
 in the right hand side of \equ{firsttrace},
 we have
 \eqb
 \dist{Y(\xi)} &\stackrel{\rm formally}{=}&
  \sum_{j=0}^\infty z^{(j)},
 \eqe
 where
 \eqb\eqa{zj}
  z^{(j)} &=&-i\Tr_{\dV}
        \left[\dfrac{r}{2a}(\xi W+W\xi)
        C^{-1}(-BC^{-1})^{j}\gamma_{d+1}\right],\ j=0,1,2,\dots.
 \eqe
 The following proposition is a consquence of the order estimates
 in \secu{order}.
 \bparr
 \prpb\prpa{zj}
 For $j>d/2$, $z^{(j)}$ is irrelevant, i.e. 
 \eqb\eqa{irrelevant}
  \lim_{a\to0} (\dist{Y(\xi)}-\sum_{j=0}^{d/2}z^{(j)}) &=& 0\,.
 \eqe
 \prpe\bparr
 \prfb
  It suffices to bound the right hand side of 
  \eqb
   \dist{Y(\xi)}-\sum_{j=0}^{d/2}z^{(j)} &=& -i\Tr_{\dV}
             \left[\dfrac{r}{2a}(\xi W+W\xi)
             (C+B)^{-1}(-BC^{-1})^{d/2+1}\gamma_{d+1}\right].
  \eqe
  The order estimates
  \equ{WO}, \equ{B+CinvO}, and \equ{BCinvO} imply
  \eqb
   \dfrac{r}{2a}(\xi W+W\xi)
   (C+B)^{-1}(-BC^{-1})^{d/2+1}\gamma_{d+1}
   &=& \O(\dfrac{a^{d/2+2}}{(1+a\dist{p}^2)^{d/2}})\,,
  \eqe
  with the help of \lemu{smoothfunction} and \lemu{sumproduct}.
  Let us take the trace on $\dV$ by means of the planewave basis
  \equ{planewavebasis}.
  Then, we have from \equ{nd1} with $q=p$
  \eqb
    |\dist{Y(\xi)}-\sum_{j=0}^{d/2}z^{(j)}|
     &\le& \const\sum_{p\in\torus^*} 
           \dfrac{a^{d/2+2}}{(1+a\dist{p}^2)^{d/2}} \nonumber\\
     &\le& \const a^2\log (1/a)\nonumber\\
     &\to& 0\ , a\to0, 
  \eqe
  where we used 
  \eqb
    \sum_{p\in\torus^*} 
     \dfrac{a^{\ell}}{(1+a\dist{p}^2)^{\ell}}
     &\le& \const 
       \left\{ \begin{array}{ll}
         a^{\ell-d/2}, & \ell>d/2, \\ 
         \log (1/a)\,, & \ell=d/2. 
       \end{array} \right. 
       \eqa{asum}
   \eqe
 \prfe\bparr
%
 \subsection{Spin Properties}
 Next we take into account the `spin properties' of operators.
 We classify operators on $\dV$
 with respect to the homogeneous degrees in $\gamma$ matrices.
 For example, 
 $C$ and $B$ have 0 and 2 homogeneous degrees, respectively.
 An operator with homogeneous degree 0
 is regarded as an operator on $\tV=\{\varphi\ :\ \torus\to\complex\}$.
 Since $\dV=\complex^{d/2}\otimes\tV$,
 the trace is factorized as:
 \eqb
  \Tr_{\dV} &=& \Tr_{\tV} \Tr_{\complex^{d/2}}.
 \eqe
 \bparr
 \prpb\prpa{zj0zd/2}
  It holds that
   \eqb
    z^{(j)} &=& 0\ ,\ j=0,1,\dots,d/2-1,            \eqa{zjto0} \\
    z^{(d/2)} &=& -i^{d/2+1}\sum_{\mu_1,\mu_2,\dots,\mu_d=1}^{d}
       \epsilon_{\mu_1\mu_2\dots\mu_{d}}
       \Tr_{\tV}[\dfrac{r}{2a}(\xi W+W\xi) \nonumber \\
        &&\hspace{30mm}
       C^{-1}X_{\mu_1\mu_2}C^{-1}X_{\mu_3\mu_4}C^{-1}\cdots
       X_{\mu_{d-1}\mu_d}C^{-1}],                   \eqa{zd/2}
   \eqe
  where $\epsilon_{\mu_1\mu_2\dots\mu_{d}}$
  denotes the totally antisymmetric tensor,
  and all the operators, having homogeneous degree 0,
  are regarded as defined on the space $\tV$.
 \prpe
 \prfb
  The operator $(\xi W+W\xi)C^{-1}(-BC^{-1})^{j}$ has
  homogeneous degree $2j$ in $\gamma$'s.
  Then, \equ{zjto0} is obvious from
  \eqb
     \Tr_{\complex^{2^{d/2}}}
     \left(\gamma_{\mu_1}\gamma_{\mu_2}\cdots\gamma_{\mu_\ell}\gamma_{d+1}
     \right) &=& 0, \
     0\le\ell<d,\ \mu_1,\mu_2,\dots,\mu_\ell=1,2,\dots,d.  \eqa{trg-}
  \eqe
  On the other hand, \equ{zd/2} follows from \equ{B} and
  \eqb
     \Tr_{\complex^{2^{d/2}}}
     \left(\gamma_{\mu_1}\gamma_{\mu_2}\cdots\gamma_{\mu_d}\gamma_{d+1}\right)
     &=& (-2i)^{d/2}\epsilon_{\mu_1\mu_2\dots\mu_{d}}, \nonumber\\
     &&\quad\mu_1,\mu_2,\dots,\mu_{d}=1,2,\dots,d.              \eqa{trg}
  \eqe 
 \prfe\bparr
 \subsection{Leading Terms}
 We extract the leading terms from \equ{zd/2}.
 Let us introduce the `normal product' : :
 that moves multiplication operators to front, e.g.
 \eqb
  :\xi W_0 + W_0\xi: &=& 2\xi W_0, \\
  :F_{\mu\nu}E_{\mu0}E_{\nu0} L_0^{-1}
   F_{\mu'\nu'}E_{\mu'0}E_{\nu'0} L_0^{-1}:
   &=& 
   F_{\mu\nu}F_{\mu'\nu'}
   E_{\mu0}E_{\nu0}L_0^{-1}E_{\mu'0}E_{\nu'0} L_0^{-1}.
 \eqe\bparr
 \prpb\prpa{barX}
  Put, for $\mu,\nu=1,2,\dots,d$,
  \eqb\eqa{barX}
   \bar X_{\mu\nu} &=&
    -iQ\left(F_{\mu\nu}E_{\mu0} E_{\nu0} L_0^{-1}
    +2ar\sum_{\rho=1}^dF_{\rho\nu}L_0^{-2}D_{\rho0}D_{\mu0}E_{\nu0}\right).
  \eqe
  and
  \eqb
   z_0 &=& -i^{d/2+1}
       \sum_{\mu_1,\mu_2,\dots,\mu_{d/2}=1}^{d}
       \sum_{\nu_1,\nu_2,\dots,\nu_{d/2}=1}^{d}
       \epsilon_{\mu_1\nu_1\dots\mu_{d/2}\nu_{d/2}} 
     \nonumber \\  &&\quad
       \Tr_{\tV}[:\dfrac{r}{a}\xi W_0
       C_0^{-1}\bar X_{\mu_1\nu_1}C_0^{-1}\bar X_{\mu_2\nu_2}C_0^{-1}\cdots
       \bar X_{\mu_{d/2}\nu_{d/2}}C_0^{-1}:],                   \eqa{z0}
  \eqe
  Then, we have
  \eqb\eqa{zd/2-z0}
   \lim_{a\to0} (z^{(d/2)}-z_{0}) &=& 0.
  \eqe
 \prpe
 \prfb
  Insert
  \eqb
   C^{-1} &=& C_0^{-1}-C_0^{-1}C_1C^{-1}
  \eqe
  into the right hand side of \equ{zd/2}
  and expand it with respect to $C_1$.
  \lemu{freeparts} and \lemu{C} imply that
  $C_0^{-1}$ has the same order as $C^{-1}$ and 
  \eqb
   C_0^{-1}C_1C^{-1} &=& 
   \O(\dfrac{1}{1+\dist{p}}\dfrac{1+a\dist{p}^2}{1+\dist{p}^2}) 
  \eqe
  is of lower order than $C^{-1}$ by the factor $1/(1+\dist{p})$.
  Then, terms containing $C_1$
  vanish in the continuum limit in a similar way as in the proof of \prpu{zj}.
  By the same reason, we can replace $W$ by $W_0$
  (see \equ{W1O}, \equ{WO}, and \equ{ndW0}).
  Furthermore, we can replace $X_{\mu\nu}$ by
  \eqb\eqa{justleadX}
   -iQ\left(F_{\mu\nu}E_{\mu0} E_{\nu0} L_0^{-1}
    +ar\sum_{\rho=1}^dF_{\rho\nu}L_0^{-2}D_{\rho0}D_{\mu0}E_{\nu0}
    -ar\sum_{\rho=1}^dF_{\rho\mu}L_0^{-2}D_{\rho0}D_{\nu0}E_{\mu0}
    \right)
  \eqe
  (see \equ{leadX}).
  Here, the second and third terms of \equ{justleadX}
  give the same contributions to $z^{(d/2)}$ 
  when \equ{leadX} is inserted into the right hand side of \equ{zd/2}.
  Then, we can replace $X_{\mu\nu}$ by $\bar X_{\mu\nu}$.
  Finally in order to replace $\xi W_0+W_0\xi$ by $2\xi W_0$
  and to take the normal product,
  it suffices to bound commutators 
  $[W_0,\xi], [C_0^{-1},F_{\mu\nu}], 
  [E_{\mu'0}E_{\nu'0}L_0^{-1},F_{\mu\nu}]$ etc.
  Those commutators are estimated using \lemu{comfreefunc}:
  By taking a commutator with a multiplication operator, 
  the order of the free part of an operator
 (with the suffix 0 like $D_{\mu0}$) 
  is reduced by the facter $1/(1+\dist{p})$, hence
  the trace containing at least one commutators 
  vanishes in the continuum limit.
 \prfe\bparr
 \subsection{Symmetry arguments}
 In view of \equ{barX} and \equ{z0},
 we see that $z_0$ is a polynomial in $r$ of order $d/2+1$ if
 we ignore the $r$-dependences of $L_0^{-1}$ and $C_0^{-1}$.
 The following proposition claims that 
 the cubic and higher order terms vanish.
 \bparr
 \prpb\prpa{z1z2}
 Put
  \eqb
   z_1 &=& -iQ^{d/2}
       \sum_{\mu_1,\mu_2,\dots,\mu_{d/2}=1}^{d}
       \sum_{\nu_1,\nu_2,\dots,\nu_{d/2}=1}^{d}
       \epsilon_{\mu_1\nu_1\dots\mu_{d/2}\nu_{d/2}} 
     \nonumber \\  && \quad
       \dfrac{r}{a}\Tr_{\tV}[:\xi W_0C_0^{-(d/2+1)}L_0^{-d/2}
       \prod_{j=1}^{d/2}\left(F_{\mu_j\nu_j}E_{\mu_j0}E_{\nu_j0}\right):],
                                                              \eqa{z1}\\
   z_2 &=& -iQ^{d/2}
       \sum_{\mu_1,\mu_2,\dots,\mu_{d/2}=1}^{d}
       \sum_{\nu_1,\nu_2,\dots,\nu_{d/2}=1}^{d}
       \epsilon_{\mu_1\nu_1\dots\mu_{d/2}\nu_{d/2}} 
     \nonumber \\  && \quad
     2 r^2\Tr_{\tV}[:\xi W_0C_0^{-(d/2+1)}L_0^{-(d/2+1)}
       \sum_{\ell=1}^{d/2}\sum_{\rho=1}^d
       F_{\rho\nu_\ell}D_{\rho0}D_{\mu_\ell0}E_{\nu_\ell0}
     \nonumber \\  && \quad\quad
       \mathop{\prod_{j=1}^{d/2}}_{j\neq\ell}
       \left(F_{\mu_j\nu_j}E_{\mu_j0}E_{\nu_j0}\right):].     \eqa{z2}
  \eqe
  Then, we have
  \eqb\eqa{z0z1z2}
   z_0 &=& z_1+z_2.
  \eqe
  Moreover, in the right hand side of \equ{z2},
  we can fix the value of $\rho$ to $\mu_\ell$, i.e.
  \eqb
   z_2 &=& -iQ^{d/2}
       \sum_{\mu_1,\mu_2,\dots,\mu_{d/2}=1}^{d}
       \sum_{\nu_1,\nu_2,\dots,\nu_{d/2}=1}^{d}
       \epsilon_{\mu_1\nu_1\dots\mu_{d/2}\nu_{d/2}} 
     \nonumber \\  && \quad
       2r^2\Tr_{\tV}[:\xi W_0C_0^{-(d/2+1)}L_0^{-(d/2+1)}
       \sum_{\ell=1}^{d/2}
       F_{\mu_\ell\nu_\ell}D_{\mu_\ell0}^2E_{\nu_\ell0}
     \nonumber \\  && \quad\quad
       \mathop{\prod_{j=1}^{d/2}}_{j\neq\ell}
       \left(F_{\mu_j\nu_j}E_{\mu_j0}E_{\nu_j0}\right):].     \eqa{z2'}
  \eqe
 \prpe
 \prfb
 Insert \equ{barX} into the summand in the right hand side of \equ{z0} 
 and expand it.
 Let us look at a term generated by the expansion 
 that contains at least two factors of the form
 $2ar\sum_{\rho=1}^dF_{\rho\nu}L_0^{-2}D_{\rho0}D_{\mu0}E_{\nu0}$
 (the second term of the right hand side of \equ{barX}).
 Such a term has the form
 \eqb
  :\cdots 2ar\sum_{\rho_\ell=1}^dF_{\rho_\ell\nu_\ell}
   L_0^{-2}D_{\rho_\ell0}D_{\mu_\ell0}E_{\nu_\ell0}
  \cdots
   2ar\sum_{\rho_{\ell'}=1}^dF_{\rho_{\ell'}\nu_{\ell'}}
   L_0^{-2}D_{\rho_{\ell'}0}D_{\mu_{\ell'}0}E_{\nu_{\ell'}0}
  \cdots:.
 \eqe
 Then, it is symmetric with respect to $\mu_\ell$ and $\mu_{\ell'}$,
 hence it vanishes when it is summed up with respect to
 $\mu_\ell$ and $\mu_{\ell'}$ 
 under the presence of the totally antisymmetric tensor.
 This gives \equ{z0z1z2}.

 Next, note that $\mu_1,\nu_1,\dots,\mu_{d/2},\nu_{d/2}$ 
 are mutually distinct because of the totally antisymmetric tensor.
 Then, the sum \equ{z2} is decomposed into partial sums
 in which $\rho$ is fixed to $\mu_j$ or $\nu_j$ for $j=1,2,\dots,d/2$.
 If $\rho=\nu_\ell$, $F_{\rho\nu_{\ell}}$ vanishes.
 It therefore suffices to consider the case 
 $\rho=\mu_j, \nu_j$ for some $j\neq\ell$.
 Assume $\rho=\mu_j$.
 In this case, the summand has the form 
 \eqb
   :\cdots F_{\mu_j\nu_\ell}D_{\mu_j0}D_{\mu_\ell0}E_{\nu_\ell0}
    \cdots F_{\mu_j\nu_j}E_{\mu_j0}E_{\nu_j0}
    \cdots:
 \eqe
 Then, it is symmetric with respect to $\nu_\ell$ and $\nu_j$,
 hence it vanishes when it is summed up with respect to
 $\nu_\ell$ and $\nu_j$.
 The term for $\rho=\nu_j$ similarly vanishes.
 Thus we have \equ{z2'}.
 \prfe\bparr
 Combining \prpu{zj}, \prpu{zj0zd/2}, \prpu{barX}, and \prpu{z1z2},
 we obtain:
 \bparr
 \prpb
  Put
  \eqb\eqa{G}
   G &=& \dfrac{r}{a}W_0C_0^{-(d/2+1)}L_0^{-(d/2+1)}
       \left(L_0+ar\sum_{\ell=1}^d\dfrac{D^2_{\ell0}}{E_{\ell0}}\right)
       \prod_{j=1}^{d}E_{j0},
  \eqe
  where we write (with a slight abuse of notaition)
  \eqb
    \dfrac{D^2_{\ell0}}{E_{\ell0}}\prod_{j=1}^{d}E_{j0}
    &=&
    D^2_{\ell0}\mathop{\prod_{j=1}^{d}}_{j\neq\ell}E_{j0}.
  \eqe
    Then we have
  \eqb\eqa{finaltrace}
   \lim_{a\to0}\left(
   \dist{Y(\xi)}+iQ^{d/2}\Tr_{\tV}(\xi\epsilon F^{d/2}G)
   \right) &=& 0,
  \eqe
  where
  \eqb
   \epsilon F^{d/2} &=& 
       \sum_{\mu_1,\mu_2,\dots,\mu_{d/2}=1}^{d}
       \sum_{\nu_1,\nu_2,\dots,\nu_{d/2}=1}^{d}
       \epsilon_{\mu_1\nu_1\dots\mu_{d/2}\nu_{d/2}} 
       \prod_{j=1}^{d/2}F_{\mu_j\nu_j}. 
  \eqe
 \prpe\bparr
 \prfb
  Rewrite the right hand sides of \equ{z1} and \equ{z2'} as:
  \eqb
   z_1 &=&
    -iQ^{d/2}\Tr_{\tV}\left(\xi\epsilon F^{d/2}
    \dfrac{r}{a}W_0C_0^{-(d/2+1)}L_0^{-d/2}\prod_{j=1}^{d}E_{j0}
    \right),                                             \eqa{z1'}\\
   z_2 &=&
    -iQ^{d/2}\Tr_{\tV}\left(\xi
       \sum_{\mu_1,\mu_2,\dots,\mu_{d/2}=1}^{d}
       \sum_{\nu_1,\nu_2,\dots,\nu_{d/2}=1}^{d}
       \epsilon_{\mu_1\nu_1\dots\mu_{d/2}\nu_{d/2}} 
       \prod_{j=1}^{d/2}F_{\mu_j\nu_j}
      \right. \nonumber\\ && \ \ \ \ \ \ \ \left.
    2r^2W_0C_0^{-(d/2+1)}L_0^{-(d/2+1)}
    \sum_{\ell=1}^{d/2}\dfrac{D^2_{\mu_\ell0}}{E_{\mu_\ell0}}
    \prod_{j=1}^{d}E_{j0}\right)                         \nonumber\\
   &=&
    -iQ^{d/2}\Tr_{\tV}\left(\xi
       \sum_{\mu_1,\mu_2,\dots,\mu_{d/2}=1}^{d}
       \sum_{\nu_1,\nu_2,\dots,\nu_{d/2}=1}^{d}
       \epsilon_{\mu_1\nu_1\dots\mu_{d/2}\nu_{d/2}} 
       \prod_{j=1}^{d/2}F_{\mu_j\nu_j}
      \right. \nonumber\\ && \ \ \ \ \ \ \ \left.
    2r^2W_0C_0^{-(d/2+1)}L_0^{-(d/2+1)}
    \dfrac12\sum_{\ell=1}^{d/2}
    (\dfrac{D^2_{\mu_\ell0}}{E_{\mu_\ell0}}
     +\dfrac{D^2_{\nu_\ell0}}{E_{\nu_\ell0}})
    \prod_{j=1}^{d}E_{j0}\right)                         \nonumber\\
   &=&
    -iQ^{d/2}\Tr_{\tV}\left(\xi\epsilon F^{d/2}
    2r^2W_0C_0^{-(d/2+1)}L_0^{-(d/2+1)}
    \dfrac12\sum_{\ell=1}^{d/2}
    \dfrac{D^2_{\ell0}}{E_{\ell0}}
    \prod_{j=1}^{d}E_{j0}\right).                        \eqa{z2''}
  \eqe
  Then, the proposition follows from
  \equ{irrelevant}, \equ{zjto0}, \equ{zd/2-z0}, \equ{z0z1z2}, \equ{z1'},
  and \equ{z2''}.
 \prfe
 \subsection{Coefficients}
 We take the trace of the left hand side of \equ{finaltrace}
 with respect to the (spinless) planewave basis $u_p, p\in\torus^*,$
 defined by \equ{planewave}.
 In the following proposition, $G(p)$ is the eigenvalue of $G$, i.e.
 $Gu_p=G(p)u_p$ (see the proof of \lemu{freeparts}).
 \prpb
  Put
  \eqb
   G(p) 
   &=& -2r^2a^d\sum_{\mu=1}^d(1-\cos ap_{\mu})
        \prod_{\lambda=1}^d\cos ap_{\lambda}               \nonumber\\
   &&\quad\quad 
        \times\dfrac{\dfrac{Ma}{r}+\sum\limits_{\nu=1}^d(1-\cos ap_{\nu})
         -\sum\limits_{\nu=1}^d\dfrac{\sin^2 ap_{\nu}}{\cos ap_{\nu}}}
        {\left[(Ma + r\sum\limits_{\nu=1}^d (1-\cos ap_{\nu}))^2
         + \sum\limits_{\nu=1}^d \sin^2 ap_{\nu}\right]^{d/2+1}}\ ,
                                                              \eqa{Gp}\\
   \kappa_a &=& \dfrac{1}{L^d}\sum_{p\in\torus^*}G(p). \eqa{kappa}
  \eqe
  Then, we have
  \eqb
   \lim_{a\to0}\left(
   \dist{Y(\xi)}
   +i\kappa_a Q^{d/2}a^d\sum_{x\in\torus}\xi(x)\epsilon F^{d/2}(x)
   \right) &=& 0,
  \eqe
  where
  \eqb
   \epsilon F^{d/2}(x) &=&
       \sum_{\mu_1,\mu_2,\dots,\mu_{d/2}=1}^{d}
       \sum_{\nu_1,\nu_2,\dots,\nu_{d/2}=1}^{d}
       \epsilon_{\mu_1\nu_1\dots\mu_{d/2}\nu_{d/2}} 
       \prod_{j=1}^{d/2}F_{\mu_j\nu_j}(x)  \ ,\ x\in\torus. 
  \eqe
 \prpe
 \prfb
  Let $\phi$ be a multiplication operator determined by a function $\varphi$
  which satisfies \equ{latter1} and \equ{latter2},
  and let $K_0$ be a free operator on $\tV$, i.e.
  an operator with $u_p, p\in\torus^*$ as the set of eigenfunctions.
  We denote the eigenvalue by $K_0(p)$, i.e. $K_0u_p=K_0(p)u_p$.
  Then we see that
  \eqb
   \Tr_{\tV}(\phi K_0)
   &=& \sum_{p\in\torus^*}(u_p, \phi K_0u_p) \nonumber\\
   &=& \sum_{p\in\torus^*} K_0(p)(u_p, \phi u_p) \nonumber\\
   &=& \sum_{p\in\torus^*} K_0(p) 
       \dfrac{a^d}{L^d}\sum_{x\in\torus}\varphi(x).
  \eqa{phiK0}\eqe
  The proposition is a consequence of \equ{finaltrace}
  and \equ{phiK0}.
 \prfe\bparr
 Now, the proof of \thmu{maintheorem} is completed
 if we confirm the following.
 \bparr
 \prpb[\protect\cite{SS}]\prpa{integralation}
  It holds that
  \eqb
   \lim_{a\to0} \kappa_a &=& \dfrac{2}{(4\pi)^{d/2}(d/2)!}.
  \eqe
 \prpe\bparr
 \prfb
  Put
  \eqb
   g(a,q) &=& a^{-d}G(q/a) \nonumber\\
    &=& -2r^2\sum_{\mu=1}^d(1-\cos q_{\mu})
        \prod_{\lambda=1}^d\cos q_{\lambda}               \nonumber\\
   &&\quad\quad 
        \times\dfrac{\dfrac{Ma}{r}+\sum\limits_{\nu=1}^d(1-\cos q_{\nu})
         -\sum\limits_{\nu=1}^d\dfrac{\sin^2 q_{\nu}}{\cos q_{\nu}}}
        {\left[(Ma + r\sum\limits_{\nu=1}^d (1-\cos q_{\nu}))^2
         + \sum\limits_{\nu=1}^d \sin^2 q_{\nu}\right]^{d/2+1}}
  \eqe
  for $a>0$, and 
  \eqb 
    g(0,q) &=& \lim_{a\downarrow 0} g(a,q)\,. 
  \eqe
  Using the funcion $g$, we write
  \eqb
   \kappa_a &=& \left(\dfrac{a}{L}\right)^d\sum_{q\in a\torus^*}g(a,q).
  \eqe
  For $q\in{\T^d}^*=[-\pi,\pi]^d$,
  let $[q]$ denote the point in $a\torus^*$ nearest to 
  $q$. $[q]$ is defined almost everywhere in ${\T^d}^*$.
  Then, it holds that
  \eqb
   \kappa_a &=& \dfrac{1}{(2\pi)^d}\int_{{\T^d}^*}dq\, g(a,[q]).
  \eqe
  Note that, for an arbitrarily small $\epsilon>0$, 
  we can choose $\delta>0$ such that
  \eqb
   \dfrac{1}{(2\pi)^d}\int_{B_\delta}dq\, |g(a,[q])| &<& \epsilon,
  \eqe
  hold for any $a\ge 0$,
  where $B_\delta=\{q\in\real^d\ |\ |q|<\delta\}$.
  Removing the small ball $B_\delta$ from the hypercube ${\T^d}^*$,
  we can take the continuum limit:
  \eqb
   \lim_{a\to0}\dfrac{1}{(2\pi)^d}
   \int_{{\T^d}^*\setminus B_\delta}dq\, g(a,[q]) 
   &=&
   \dfrac{1}{(2\pi)^d}
   \int_{{\T^d}^*\setminus B_\delta}dq\, g(0,q).
  \eqe
  Then, it holds that
  \eqb
   \lim_{a\to0}\kappa_a
   &=& \dfrac{1}{(2\pi)^d}\int_{{\T^d}^*}dq\, g(0,q).
  \eqe
  It therefore suffices to show
  \eqb\eqa{Omega}
  \int_{{\T^d}^*}dq\, g(0,q)
   &=& \dfrac{2}{d}\Omega_d,
  \eqe
  where $\Omega_d$ stands for the area of the $d$-dimensional unit sphere,
  i.e. $\Omega_d=(2\pi)^{d/2}/(d-2)!!$.
  This fact is proved for $d=4$ in \cite{KS} and 
  for all $d$ and all $r$ in \cite{SS}.\footnote{%
  To be strict, our conclusion is consistent with that of \cite{SS}, if
 we include $\prod_{\mu} \epsilon_{\mu}$ and replace $s^d-1$ by
 $s^{d-1}$ in the integrand of (A20) in \cite{SS}.}
 \prfe
 
 \bparr\appendix
 We list up the operators that are (globally) used in this paper.
 $$
  T_{\mu0}u(x)=u(x+ae_\mu)                    \eqno\equ{freetranslation}
 $$
 $$
  T_{\mu}=U_{\mu}T_{\mu0}                     \eqno\equ{covtranslation}
 $$
 $$
  D_\mu=\dfrac{1}{2a}(T_\mu-T_\mu^*)          \eqno\equ{Dmu}
 $$
 $$
  \Slash D = \sum_{\mu=1}^{d}\gamma_\mu D_\mu \eqno\equ{Diracoperator}
 $$
 $$
  W=-\sum_{\mu=1}^{d}(2I-T_\mu-T_\mu^*)       \eqno\equ{Wilson}
 $$
 $$
   L = MI - \dfrac{r}{2a}W                    \eqno\equ{L}
 $$
 $$
   X_{\mu\nu} = D_\mu L^{-1}D_\nu-D_\nu L^{-1}D_\mu  \eqno\equ{X}
 $$
 $$
   B = \dfrac12 \sum_{\mu,\nu=1}^{d}\gamma_{\mu}\gamma_{\nu}X_{\mu\nu}  
                                              \eqno\equ{B}
 $$
 $$
   C = L - \sum_{\mu=1}^{d} D_\mu L^{-1}D_\mu \eqno\equ{C}
 $$
 $$
  E_{\mu0} = \dfrac12(T_{\mu0}+T^*_{\mu0})    \eqno\equ{Emu0}
 $$
 

\end{document}